\documentclass[final,3p,times]{elsarticle}
\usepackage{amssymb}
\usepackage{amsmath}
\usepackage[version=4]{mhchem}
\usepackage{bm}
\usepackage{algorithm}
\usepackage{algpseudocode}
\usepackage{mathtools}
\usepackage{lineno,hyperref}
\usepackage{rotating}
\usepackage{enumitem}
\usepackage{parskip}

\usepackage{subcaption}
\usepackage{makecell}%

\journal{TBD}

\biboptions{sort&compress}
\usepackage[dvipsnames]{xcolor}

\usepackage{lineno}

\usepackage{tikz}
\usepackage{etoolbox}
\newrobustcmd*{\myVtriangle}[2]{\tikz{\filldraw[draw=#1,fill=#2] (0cm,0.2cm) --
(0.2cm,0.2cm) -- (0.1cm,0cm) -- (0cm,0.2cm);}}
\newrobustcmd*{\mythickVtriangle}[2]{\tikz{\filldraw[line width=0.3mm,draw=#1,fill=#2] (0cm,0.2cm) --
(0.2cm,0.2cm) -- (0.1cm,0cm) -- (0cm,0.2cm);}}
\newrobustcmd*{\mythickErrorVtriangle}[2]{\tikz{\filldraw[line width=0.3mm,draw=#1,fill=#2] (-0.05cm,0.05cm) --
(0.05cm,0.05cm) -- (0cm,-0.05cm) -- (-0.05cm,0.05cm);  \draw[draw=#1] (0.0cm, -0.12cm) -- (0.0cm, 0.12cm) ; \draw[draw=#1] (-0.06cm, 0.12cm) -- (0.06cm, 0.12cm); \draw[draw=#1] (-0.06cm, -0.12cm) -- (0.06cm, -0.12cm)    }}
\newrobustcmd*{\mytriangle}[2]{\tikz{\filldraw[draw=#1,fill=#2] (0.0cm,0.0cm) --
(0.2cm,0cm) -- (0.1cm,0.2cm) -- (0cm,0cm);}}
\newrobustcmd*{\mysquare}[2]{\tikz{\draw[draw=#1,fill=#2] (0cm,0cm)
rectangle (0.2cm,0.2cm)}}
\newrobustcmd*{\mythicktriangle}[2]{\tikz{\filldraw[line width=0.3mm,draw=#1,fill=#2] (0.0cm,0cm) --
(0.2cm,0cm) -- (0.1cm,0.2cm) -- (0.0cm,0cm);}}
\newrobustcmd*{\mythicksquare}[2]{\tikz{\draw[line width=0.3mm,draw=#1,fill=#2] (0cm,0cm)
rectangle (0.2cm,0.2cm)}}
\newrobustcmd*{\mybarredtriangle}[2]{\tikz{\draw[draw=#1,fill=#2] (0,0) --
(0.2cm,0) -- (0.1cm,0.2cm) -- (0cm,0cm); \draw[draw=#1] (-0.1cm, 0.07cm) -- (0.3cm, 0.07cm)}}
\newrobustcmd*{\mythickbarredtriangle}[2]{\tikz{\draw[line width=0.3mm,draw=#1,fill=#2] (0,0) --
(0.2cm,0) -- (0.1cm,0.2cm) -- (0cm,0cm); \draw[draw=#1] (-0.1cm, 0.07cm) -- (0.3cm, 0.07cm)}}
\newrobustcmd*{\mybarredsquare}[2]{\tikz{\draw[draw=#1,fill=#2] (0,0)
rectangle (0.2cm,0.2cm); \draw[draw=#1] (-0.1cm, 0.1cm) -- (0.3cm, 0.1cm)}}
\newrobustcmd*{\mythickbarredsquare}[2]{\tikz{\draw[line width=0.3mm,draw=#1,fill=#2] (0,0)
rectangle (0.2cm,0.2cm); \draw[draw=#1] (-0.1cm, 0.1cm) -- (0.3cm, 0.1cm)}}
\newrobustcmd*{\mybarredcircle}[2]{\tikz{\draw[draw=#1,fill=#2] (0,0)
circle (0.1cm); \draw[draw=#1] (-0.2cm, 0.0cm) -- (0.2cm, 0.0cm)}}
\newrobustcmd*{\mythickbarredcircle}[2]{\tikz{\draw[line width=0.3mm,draw=#1,fill=#2] (0,0)
circle (0.1cm); \draw[draw=#1] (-0.2cm, 0.0cm) -- (0.2cm, 0.0cm)}}
\newrobustcmd*{\mythickErrorcircle}[2]{\tikz{\draw[line width=0.3mm,draw=#1,fill=#2] (0,0)
circle (0.06cm); \draw[draw=#1] (0.0cm, -0.12cm) -- (0.0cm, 0.12cm) ;   \draw[draw=#1] (-0.06cm, 0.12cm) -- (0.06cm, 0.12cm); \draw[draw=#1] (-0.06cm, -0.12cm) -- (0.06cm, -0.12cm)    }}
\newrobustcmd*{\mythickErrortriangle}[2]{\tikz{\draw[line width=0.3mm,draw=#1,fill=#2] (0.0cm,-0.05cm) --
(0.15cm,-0.05cm) -- (0.075cm,0.1cm) -- (0.0cm,-0.05cm);  \draw[draw=#1] (0.075cm, 0.1cm) -- (0.075cm, 0.15cm); \draw[draw=#1] (0.075cm, -0.05cm) -- (0.075cm, -0.1cm); \draw[draw=#1] (0.0cm, 0.15cm) -- (0.15cm, 0.15cm); \draw[draw=#1] (0.0cm, -0.1cm) -- (0.15cm, -0.1cm); } }
\newrobustcmd*{\mydashedline}[1]{\tikz{\draw[draw=#1] (-0.2cm, 0.2cm) -- (-0.1cm, 0.2cm); \draw[draw=#1] (-0.0cm, 0.2cm) -- (0.1cm, 0.2cm)}}
\newrobustcmd*{\mythickcross}[1]{\tikz{\draw[line width=0.3mm,draw=#1] (0,0) --
(0.2cm,0); \draw[line width=0.3mm,draw=#1] (0.1cm,-0.1cm) -- (0.1cm,0.1cm);}}
\newrobustcmd*{\mybarredcross}[1]{\tikz{\draw[line width=0.3mm,draw=#1] (0,0) --
(0.2cm,0); \draw[line width=0.3mm,draw=#1] (0.1cm,-0.1cm) -- (0.1cm,0.1cm); \draw[draw=#1] (-0.1cm,0) -- (0.3cm,0);}}
\newrobustcmd*{\myline}[1]{\tikz{\draw[draw=#1] (-0.15cm, 0.1cm) -- (0.15cm, 0.1cm);\draw[line width=0.3mm,draw=#1] (-0.0cm, 0.0cm);}}
\newrobustcmd*{\mythickline}[1]{\tikz{\draw[line width=0.3mm,draw=#1] (-0.15cm, 0.1cm) -- (0.15cm, 0.1cm);\draw[line width=0.3mm,draw=#1] (-0.0cm, 0.0cm);}}
\newrobustcmd*{\mythickdashedline}[1]{\tikz{\draw[line width=0.3mm,draw=#1] (-0.2, 0.1cm) -- (-0.1cm, 0.1cm); \draw[line width=0.3mm,draw=#1] (-0.0cm, 0.1cm) -- (0.1cm, 0.1cm); \draw[line width=0.3mm,draw=#1] (-0.0cm, 0.0cm);}}
\newrobustcmd*{\mythickdasheddottedline}[1]{\tikz{\draw[line width=0.3mm,draw=#1] (-0.22, 0.1cm) -- (-0.13cm, 0.1cm); \draw[line width=0.3mm,draw=#1] (-0.085cm, 0.1cm) -- (-0.055cm, 0.1cm); \draw[line width=0.3mm,draw=#1] (-0.01cm, 0.1cm) -- (0.08cm, 0.1cm); \draw[line width=0.3mm,draw=#1] (-0.0cm, 0.0cm);}}
\newrobustcmd*{\mycircle}[2]{\tikz{\draw[draw=#1,fill=#2] (0,0)
circle (0.1cm);}}
\newrobustcmd*{\mythickcircle}[2]{\tikz{\draw[line width=0.3mm,draw=#1,fill=#2] (0,0)
circle (0.1cm);}}
\newrobustcmd*{\mydot}[1]{\tikz{\draw[line width=0.3mm,draw=#1] (0,0)
circle (0.025cm);}}
\makeatletter
\newcommand{\pushright}[1]{\ifmeasuring@#1\else\omit\hfill$\displaystyle#1$\fi\ignorespaces}
\newcommand{\pushleft}[1]{\ifmeasuring@#1\else\omit$\displaystyle#1$\hfill\fi\ignorespaces}
\makeatother

\begin{document}

\begin{frontmatter}

\title{Neural posterior estimation for scalable and \\accurate  inverse parameter inference in Li-ion batteries}

\author[NRL]{Malik Hassanaly}
\author[MTES-NRL]{Corey R.~Randall}
\author[MTES-NRL]{Peter J.~Weddle}
\author[MTES-NRL]{\\Paul J.~Gasper}
\author[NRL]{Conlain Kelly}
\author[INL]{Tanvir R. Tanim}
%\author[INL]{Eric J.~Dufek}
\author[MTES-NRL]{Kandler Smith}

%ORCID
\address[NRL]{Computational Science Center, National Laboratory of the Rockies, Golden, CO 80401}
\address[MTES-NRL]{Energy Conversion and Storage Systems Center, National Laboratory of the Rockies, Golden, CO 80401}
\address[INL]{Energy Storage Research and Analysis Department, Idaho National Laboratory, Idaho Falls, ID 83415}

\begin{abstract}

Diagnosing the internal state of Li-ion batteries is critical for battery research, operation of real-world systems, and prognostic evaluation of remaining lifetime. By using physics-based models to perform probabilistic parameter estimation via Bayesian calibration, diagnostics can account for the uncertainty due to model fitness, data noise, and the observability of any given parameter. However, Bayesian calibration in Li-ion batteries using electrochemical data is computationally intensive even when using a fast surrogate in place of physics-based models, requiring many thousands of model evaluations. A fully amortized alternative is neural posterior estimation (NPE). NPE shifts the computational burden from the parameter estimation step to data generation and model training, reducing the parameter estimation time from minutes to milliseconds, enabling real-time applications. The present work shows that NPE calibrates parameters equally or more accurately than Bayesian calibration, and we demonstrate that the higher computational costs for data generation are tractable even in high-dimensional cases (ranging from 6 to 27 estimated parameters), but the NPE method can lead to higher voltage prediction errors. The NPE method also offers several interpretability advantages over Bayesian calibration, such as local parameter sensitivity to specific regions of the voltage curve. The NPE method is demonstrated using an experimental fast charge dataset, with parameter estimates validated against measurements of loss of lithium inventory and loss of active material.  The implementation is made available in a companion repository (\hyperlink{https://github.com/NatLabRockies/BatFIT}{https://github.com/NatLabRockies/BatFIT}).

%250 words max

\end{abstract}

\end{frontmatter}

%\linenumbers
%\tableofcontents
\section{Introduction}
\label{sec:intro}

\subsection{Motivation}

Li-ion batteries are essential for modern, rechargeable device electrification, ranging in scale from watches and laptops to electrified transportation and utility scale energy storage.  In these devices, it is important to use diagnostic techniques to determine the battery's internal state, for example, the ``state-of-charge'' (SOC) and ``state-of-health'' (SOH).  A battery's SOC (i.e., the remaining capacity available in the cell) is typically inferred from voltage and overpotential estimates~\cite{weddle2023battery}. Similarly, a battery's SOH (i.e., the degradation of physical processes within the battery as the cell ages) is also an important quantity to continuously monitor~\cite{weddle2023battery,kim2022rapid,duan2022parameter}.  However, internal processes at stake during battery aging depend on its cycling history, are chemistry-specific, and cell-specific, resulting in a significantly harder diagnostic~\cite{li2022data}.  Typical SOH diagnostic approaches require the cell to be operated at relatively slow cycling rates (e.g., C/10) and use beginning-of-life thermodynamic measurements to determine aging modes such as loss-of-lithium-inventory (LLI) and loss-of-active-material (LAM) of either electrode. In the present work,  novel parameter inference (i.e., inferring the internal non-observable parameters) techniques are introduced to determine the battery's SOH from high-rate ($\geq$4C) charging and moderate-rate (C/2) discharging. %Importantly, both the Bayesian calibration and Neural Posterior Estimation (NPE) technique used for SOH diagnostics not only provide estimated internal parameters from high-rate cycling, but also provide statistics on how confident the technique is at determining each internal parameter.  
Since SOH has several meanings in the battery community (e.g., can be based on capacity fade, energy fade, power fade, including associated aging-modes and mechanisms~\cite{weddle2023battery}), in the present manuscript, ``parameter inference'' is used instead.  Throughout the manuscript parameter inference indicates that the diagnostic techniques used herein determine internal parameters for a physics-based model.

Besides SOH diagnostics, parameter inference for Li-ion battery models is also essential for a broad range of engineering tasks, because it links physical properties to observed battery behavior~\cite{andersson2022parametrization}. In particular, parameter inference typically allows for creating a digital twin, which can be used for predictive modeling of aging behavior with respect to cycling~\cite{dufek2022developing}. In both predictive modeling and SOH diagnostics, parameter inference informs practical decision-making regarding how the battery will be operated, maintained, or rejuvenated. Thus, parameter inference needs to be accurate enough to avoid erroneous, and possibly costly decisions~\cite{guittet2024levelized, reniers2021unlocking}. If implemented over a large fleet of devices, or if used to extract aging trends~\cite{zhang2026discovery}, parameter inference also needs to be rapid, so it can be performed at scale or onboard.  

In what follows, the physical properties or parameters to be identified are noted as $\theta \in \mathbb{R}^{N_{\theta}}$, where $N_\theta$ is the number of internal parameters being determined. In previous studies, $N_{\theta}$ ranges between 2~\cite{hassanaly2024pinn2} and 44~\cite{reddy2019accelerating}. The battery behavior observations are noted as $x \in \mathbb{R}^{N_x}$, where the number of observable signals $N_x$ depends on the experimental observations acquired for the battery. Typically $N_{\theta} \ll N_{x}$ and usually, $x$ denotes voltage measurements over time. Parameter inference is an \textit{inverse problem} that aims to transform $x$ into $\theta$. This transformation is achieved by using a physics-based model $\mathcal{M}$~\cite{santhanagopalan2006review} that can transform $\theta$ into synthetic observations. For example, $\mathcal{M}$ could be a single particle model \cite{santhanagopalan2006review} or a pseudo-2D model ~\cite{doyle1993modeling,fuller1994relaxation}. The mapping from $\theta$ to synthetic observations is called the \textit{forward model}.

\subsection{Related work}

The most common approach for parameter inference (see Andersson et al.~\cite{andersson2022parametrization} and references therein) is to formulate the following optimization problem
\begin{equation}
    \operatorname{argmin}_{\theta} || \mathcal{M}(\theta) - x ||,
\end{equation}
where $||.||$ denotes a norm (usually a mean square error or a mean absolute error) that quantifies how close the predicted observations $\mathcal{M}(\theta)$ are to the true observations $x$. This approach is, however, a) slow, requiring several forward model evaluations; and b) ill-posed as many parameter sets $\theta$ may explain the same observation $x$~\cite{ramadesigan2011parameter,yu2025state,andersson2022parametrization}.

To resolve the computational speed issues, machine learning (ML) can be leveraged to either replace the forward model with a faster surrogate~\cite{hassanaly2024pinn1} or use a neural net to directly predict $\theta$~\cite{li2024deep,ko2024using,lenzi2023neural}. The latter application of ML is natural here, as solving the inverse problem requires identifying patterns in the voltage observation $x$ that reveal the value of $\theta$, a task where neural networks excel. 

To resolve the ill-posedness issues, three main approaches have emerged in the Li-ion battery modeling community. First, the ill-posedness can be resolved by choosing observations that are more informative about $\theta$~\cite{brendel2025parametrized,roman2022design}, but this requires performing specific measurements that may not reflect real-world use and cannot help when performing analysis on historical data where these measurements were not recorded. Second, efforts are dedicated to identifying which parameters $\theta_i$ are the most sensitive to changes in voltage observations $x$. Then, only the most sensitive parameters are identified initially, while the least sensitive are identified later~\cite{li2022data} or simply, not identified. The set of sensitive parameters is usually obtained with correlation analysis~\cite{li2024deep} or sensitivity analysis, either globally~\cite{ko2024using} or locally in parameter space~\cite{wang2023parameter}. The benefit of this two-stage approach is to avoid overfitting voltage observations by only identifying parameters that can be identified with high sensitivity. However, the two-stage approach assumes that insensitive parameters are less useful to identify as compared to sensitive parameters. Additionally, global and local sensitivity methods assume that sensitivity is homogeneous, or varies smoothly throughout parameter space. These assumptions are not always valid and are computationally expensive to assess.
Third, probabilistic calibration approaches that approximate the posterior distribution $p(\theta|x)$ (that is the distribution that describe the probability of observing $\theta$ given the observation $x$) have been demonstrated~\cite{nascimento2023framework,hassanaly2024pinn2, kim2023bayesian,aitio2020bayesian, bills2023massively}. Deploying these methods requires Bayesian updates performed by using the Bayes theorem 
\begin{equation}
    \label{eq:bayesth}
    p(\theta|x) = \frac{p(x|\theta) p(\theta)}{p(\theta,x)},
\end{equation}
where $p(x|\theta)$ is the likelihood probability, $p(\theta)$ is the prior probability and $p(\theta,x)$ is the evidence. These methods handle the ill-posedness by making the predictions probabilistic without needing to classify parameters by their sensitivity. However, probabilistic Bayesian calibration is computationally expensive~\cite{bills2023massively}. It usually makes use of Markov Chain Monte-Carlo (MCMC), requiring repeated evaluations of the likelihood. Hereafter, Bayesian calibration via MCMC is simply referred to as \textit{Bayesian calibration}. In Bayesian calibration, the likelihood evaluations require forward model evaluations~$\mathcal{M}$. Depending on the nature of the forward model and the cycling protocol, evaluating $\mathcal{M}$ can be expensive (ranges from seconds to minutes per forward model evaluation). Recently, it has been shown that one way to reduce the cost of Bayesian calibration is to construct a fast surrogate for $\mathcal{M}$~\cite{hassanaly2024pinn1}. However, even using a fast surrogate model to approximate the physics-based model $\mathcal{M}$ can take several minutes to perform Bayesian calibration for one voltage time-series~\cite{hassanaly2024pinn2}. This makes onboard parameter calibration and calibration for many devices expensive. 

%We note that soon after our first demonstration of NPE for battery parameter inference \cite{hassanaly2025surrogate}, a separate analysis has utilized NPE to inform capacity fade prognostics \cite{zhang2026discovery}.  was  for capacity fade prediction has utilized NPE study developed in parallel has utilized NPE 
A recent analysis by Zhang et al.~\cite{zhang2026discovery} demonstrated how parameters $\theta$ inferred with Neural posterior estimation (NPE) can serve as features that inform capacity fade prognostics in Li-ion batteries. In their work, the focus was on prognostics and the authors only noted that NPE reduced inference costs when compared to MCMC, which enabled deployment over several datasets, but neglected to discuss parameter or voltage estimation accuracy. In contrast, the present work focuses on systematic evaluation of the advantages and drawbacks of NPE over other methods.

\subsection{Contributions of this work}

In this work, we propose to replace the Bayesian calibration with NPE. The contributions of this work are that:
\begin{enumerate}[noitemsep,topsep=0em]
    \item We show with synthetic data that NPE is as accurate as Bayesian calibration for parameter estimation, while reducing inference costs from minutes to milliseconds.
    \item We show that NPE offers several interpretability advantages over Bayesian calibration, for instance, local parameter sensitivity to specific regions of the voltage curve.
    \item We show with experimental data that NPE can accurately predict typical battery aging modes including loss of lithium inventory and loss of active material.
    \item We show how physics-based constraints can be enforced on the parameters inferred with NPE.
    \item We show that data generation for training the NPE model is computationally tractable even for a very large number of parameters to identify ($N_{\theta}$ up to 27).
\end{enumerate}

The manuscript is organized as follows: Sec.~\ref{sec:methods} presents the NPE method and provides theoretical background; Sec.~\ref{sec:comparison} compares the NPE method to Bayesian calibration using synthetic data; Sec.~\ref{sec:application_exp} shows how NPE performs on experimental data and illustrates how the NPE method can be deployed in practice; Sec.~\ref{sec:tractability} discusses the data requirements for the NPE method when dealing with large parameter spaces; conclusions are provided in Sec.~\ref{sec:conclusions}.

\section{Methods}
\label{sec:methods}

The present section describes the methods employed for the Bayesian calibration and NPE method.  The parameters identified from the two methods are compared in Sec.~\ref{sec:comparison}-\ref{sec:tractability}. First, theoretical justification is provided for why NPE is appropriate. Second, the implementation of the methods is discussed. 

\subsection{Approximation of Bayesian calibration}

The inverse model problem targeted here falls in the category of simulation-based inference, where synthetic simulation data is used to explain experimental observations, and formulate physics-based conclusions. Depending on the simulation cost and the availability of data, there are several alternatives to Bayesian calibration that can be envisioned~\cite{cranmer2020frontier,deistler2025simulation}.

\subsubsection{Variational approximation}

In the Bayesian calibration approach, $p(\theta|x)$ is described by samples drawn from that distribution and that are obtained with MCMC. These samples are referred to as \textit{posterior samples}. Drawing posterior samples with MCMC requires building a sequence (a chain) of samples that eventually converges to independent samples of $p(\theta|x)$. Drawing each element of the chain can require tens of likelihood evaluations~\cite{hoffman2014no} and typically thousands of samples are needed.  In theory, MCMC can approximate arbitrarily complex distributions given enough chains and long enough chains, and is usually considered the gold standard of posterior sampling~\cite{nemeth2021stochastic}. Instead of using MCMC, one can approximate $p(\theta|x)$ with another probability density function (PDF) $q_{\phi}(\theta|x)$ where $\phi$ parameterizes the posterior. This is the \textit{variational approximation}. A drawback of the variational approach is that one needs to first assume a functional form of the posterior PDF $q_{\phi}(\theta|x)$, which inherently runs the risk of assuming an erroneous form (discussed further in Sec.~\ref{sec:comparison}). The key advantage to the variational approach is that drawing long chains of samples can be replaced with an optimization problem represented mathematically as 
\begin{equation}
    \label{eq:var_approx}
    \operatorname{argmin}_{\phi} d\bigg(p(\theta|x),  q_{\phi}(\theta|x)\bigg),
\end{equation}
where $d(.)$ is a PDF distance.

\subsubsection{Neural posterior estimation (NPE)}

There exist multiple PDF distances that can be used in Eq.~\ref{eq:var_approx} and among them, the Kullback-Leibler divergence (KLD)~\cite{kullback1951information} is classically employed to make $d(.)$ tractable. The KLD is positive but is not symmetric, meaning that 
\begin{align}
    \operatorname{KLD}(q_{\phi}(\theta|x) || p(\theta|x)) &\neq \operatorname{KLD}(p(\theta|x) || q_{\phi}(\theta|x)),\quad {\rm i.e.,}\\
    \int q_\theta(\theta|x) \log\left(\frac{q_{\phi}(\theta|x)}{p(\theta|x)}\right) {\rm d}\theta &\neq \int p(\theta|x) \log\left(\frac{p(\theta|x)}{q_{\phi}(\theta|x)}\right) {\rm d}\theta.
\end{align}

Using $\operatorname{KLD}(q_{\phi}(\theta|x) || p(\theta|x))$ leads to Variational Inference methods~\cite{blei2017variational}, which still require a likelihood. Though Hassanaly et al.~\cite{hassanaly2024pinn1} showed that fast likelihood sampling was possible with ML-based surrogates, they also showed that the size of the ML-surrogate was heavily influenced by the cycling protocol. In practice, the cycling protocol may not be known ahead of time, which may make it difficult to appropriately size the surrogate model. Additionally, not all parts of the cycling protocol are informative about the parameters $\theta$. Thus, while using fast physics-based model surrogates is possible, building these surrogates can easily become an unnecessarily arduous task in the context of Li-ion batteries, preventing broad applicability.

Alternatively, using $\operatorname{KLD}( p(\theta|x)|| q_{\phi}(\theta|x) ) = \mathbb{E}_{p(\theta|x)} (\operatorname{log} p(\theta|x) - \operatorname{log} q_{\phi}(\theta|x))$ does not require likelihood evaluations if many pairs of $\theta$ and $x$ can be made available, which is the case here. Since $\operatorname{log} p(\theta|x)$ does not depend on $\phi$, one can solve Eq.~\ref{eq:var_approx} by minimizing $\mathbb{E}_{p(\theta|x)} \bigg(- \operatorname{log} q_{\phi}(\theta|x)\bigg)$. Additionally, ensuring that the distance is minimized for any observation $x$ (applying an expectation $\mathbb{E}_{p(x)}(.)$) leads to the loss function
\begin{equation}
    \label{eq:loss}
    \mathcal{L} =  \mathbb{E}_{p(x,\theta)} \bigg(- \operatorname{log} q_{\phi}(\theta|x)\bigg).
\end{equation}
This approach is the \textit{Neural Posterior Estimation} approach and can be handled with a neural network trained with Eq.~\ref{eq:loss} as the loss function~\cite{papamakarios2016fast,deistler2025simulation}. At this point, it is useful to compare Bayesian calibration and the NPE method: 
\begin{itemize}[noitemsep,topsep=0em]
    \item In Bayesian calibration, the parameter priors are explicitly included in Eq.~\ref{eq:bayesth}, affecting how samples of the MCMC chains are drawn. In NPE, the priors are also included explicitly as they are used to sample pairs of (x, $\theta$) when assembling the dataset, or by including a proposal prior correction term~\cite{papamakarios2016fast}. In the present work, the parameters $\theta$ are sampled according to the priors (i.e., proposal prior correction is not used). The sample-based treatment of the prior is leveraged in Sec.~\ref{sec:results} to enforce constraints on the inferred parameters.
    \item In Bayesian calibration, an explicit functional form for the likelihood $p(x|\theta)$ is chosen, usually as a multivariate normal~\cite{braman2013bayesian,hassanaly2025bayesian}. In NPE, there is no likelihood function.
    \item In Bayesian calibration, the posterior $p(\theta|x)$ is represented by samples drawn iteratively with MCMC. In NPE, the posterior is represented by parameters predicted by a model~\cite{deistler2025simulation}.
    \item NPE is a form of fully amortized probability inference, where computational cost is invested once to generate the training data and train the model, but inference given a new observation $x$ only requires one or a few forward passes~\cite{wildberger2023flow}. In other words, the initial costs for the NPE method are amortized over many inference tasks. Bayesian calibration can only be semi-amortized if a surrogate is built to evaluate the likelihood~\cite{hassanaly2024pinn2} but still requires building a chain of samples for every new observation $x_i$.
\end{itemize}

\subsection{Li-ion battery implementation}

In this section, the Bayesian calibration and NPE implementations are described.

\subsubsection{Dataset assembly}

The dataset $\mathcal{D}$ needed to train an NPE must contain pairs of $(\theta,x)$ where $\theta$ is sampled according to its prior distribution, i.e., $\mathcal{D}=\{ (\theta_i,x_i), x_i =\mathcal{M}(\theta_i), ~ \theta_i \sim p(\theta), ~i \in [1,N_{\mathcal{D}}]\}$. For NPE to be accurate, the number of data points $N_{\mathcal{D}}$ needs to be large enough to span the support of the parameter priors. The higher the number of parameters inferred $N_{\theta}$ and the wider the prior supports, the more pairs $(\theta_i,x_i)$ are needed. The specific data requirements as a function of $N_{\theta}$ are further described in Sec.~\ref{sec:tractability}. For NPE, data generation is embarrassingly parallel. In the companion repository\footnote{\hyperlink{https://github.com/NatLabRockies/BatFIT}{https://github.com/NatLabRockies/BatFIT}}, we show a practical example of how to generate training data in parallel via message passing interface (MPI) parallelization.

In practice, experimental data can be subject to noise, which is not included in the result of a physics-based model. The noise magnitude can either be estimated based on information provided by the cycler manufacturer~\cite{hassanaly2024pinn1}, or approximated from the irreducible error between an optimal estimator and the experimental data~\cite{hassanaly2022adversarial}. Once an observation noise model is defined, one can add the noise to the training data to emulate experimental observations. Here, noise is randomly sampled and added to $\mathcal{D}$ at training time. In other words, $\forall \theta,~\mathcal{M}(\theta) = \mathcal{\widehat{M}}(\theta) + \eta$, where $\widehat{\mathcal{M}}$ is the physics-based model, and $\eta \sim p(\eta)$ where $\eta$ is the noise and $p(\eta)$ is the noise model. Another approach to incorporating noise is to augment the dataset $\mathcal{D}$ with replicates subject to different noise samples, but this approach results in larger datasets.

The physics-based simulations are conducted with an in-house solver, BATMODS-lite\footnote{\hyperlink{https://github.com/NatLabRockies/batmods-lite}{https://github.com/NatLabRockies/batmods-lite}}~\cite{randall2025bmlite}. Additional simulation and solver details are provided in \ref{app:bmlite}.

\subsubsection{Datasets}
\label{sec:datasets}

Two datasets are assembled hereafter. The first dataset is called the \textit{Comparison dataset} and is used throughout Sec.~\ref{sec:comparison} to compare Bayesian calibration and NPE results. The second dataset is called the \textit{XCEL dataset}~\cite{chen2022battery} and is used to validate NPE against experimental data in Sec.~\ref{sec:application_exp}. Additional datasets to assess the tractability of the method for high-dimensional parameter spaces are described in Sec.~\ref{sec:tractability}. 

In both datasets, the battery chemistry being simulated is a Li-ion battery with a graphite negative electrode, \ce{Li$_x$Ni_{0.5}Mn_{0.3}Co_{0.2}O_2} positive electrode, and a 1.2~M \ce{LiPF_6} salt in a 1:1 ethylene carbonate:ethyl methyl carbonate (EC:EMC) solvent; the battery chemistry determines material-specific properties, like open-circuit potential with respect to lithium concentration, but is not a variable that changes during battery operation and thus does not impact the parameter estimation.  The observation $x$ is the voltage time-series and the cycling protocols are constant-current charge and constant-current discharge. Under more complex cycling protocols, the NPE approach is still applicable (not shown here for the sake of concision) but Bayesian calibration with a likelihood surrogate can be difficult to conduct~\cite{hassanaly2024pinn1}. Using constant-current cycling achieves two goals: 1) it provides a meaningful comparison between Bayesian calibration and NPE and 2) it allows applying the NPE to the experimental data used in Sec.~\ref{sec:application_exp}. A key issue with constant-current charge and discharge is that the time spanned by the measurements varies depending on the cell properties. To address varying timespans, $x$ contains two one-dimensional arrays, where one gives the time at which voltage is measured and the other gives the voltage recorded. The voltage measurements are $N_{\rm t} = 256$ points equally spaced over time, so $N_x = N_{\rm t} \times 2 \times n_{\rm obs}$ where $n_{\rm obs}$ is the number of charge/discharge observation time series. Similar to Ref.~\cite{hassanaly2024pinn2}, the parameters calibrated are unitless efficiency factors that scale reference parameter values, allowing for a homogenized parameter scale. We note $X = X' X_{\rm ref}$, where $X$ is the scaled parameter value, $X'$ is the unitless efficiency factor, and $X_{\rm ref}$ is a reference value. 
The reference values are provided in \ref{app:bol}. For all datasets, the noise model is $\eta \sim \mathcal{U}(-1.44~\rm{mV}, 1.44~\rm{mV})$. The effect of the noise model is further discussed in \ref{app:noise}.

\vspace{0.5mm}
\noindent \underline{Comparison dataset}: The physics-based model used here is the single particle model (SPM) \cite{santhanagopalan2006review} during $\rm{C}/2$ discharge ($n_{\rm obs} = 1$). Another dataset is built in the same way for a $4\rm{C}$ charge. In both data sets, $N_{\theta}=6$ and the priors of the efficiency factors are independent uniform distributions with bounds given by Tab.~\ref{tab:priorcompds}.
%and the calibrated efficiency factors are chosen so as to be able to estimate loss of lithium inventory, and loss of active material in the positive electrode. 

\begin{table}[h!]
\centering
\begin{tabular}{|c | c | c|} 
 \hline
 Scaled parameter & Discharge ranges & Charge Ranges \\ [0.5ex] 
 \hline
 Anode exchange current density $i_{0,\rm{a}}'$ & [0.1, 4.0] & Same as discharge \\ 
 Solid cathode diffusivity $D_{\rm s,c}'$ & [0.2, 10.0] & Same as discharge \\
 Initial anode intercalation fraction $x_{0,\rm{a}}'$ & [0.6, 1.077] & [0.3, 2.0] \\
 Initial cathode intercalation fraction  $x_{0,\rm{c}}'$ & [0.88, 1.6] & [0.78, 1.08] \\
 Cathode exchange current density $i_{0,\rm{c}}'$ & [0.1, 1.6] & Same as discharge \\ 
 Solid cathode active material volume fraction $\varepsilon_{\rm{s, c, AM}}'$  & [0.7, 1.0] & Same as discharge \\[1ex] 
 \hline
\end{tabular}
\caption{Ranges of the uniform priors used for the parameters calibrated with the Comparison dataset and the XCEL dataset. The parameters calibrated are unitless efficiency factors.}
\label{tab:priorcompds}
\end{table}

The training charge and discharge datasets contain 90,000 pairs of $(\theta, x)$, requiring a 15~min calculation using 208 Intel Xeon Sapphire Rapids CPU cores. A held-out test dataset of 10,000 pairs of $(\theta, x)$ is used for hyperparameter optimization. The accuracy metrics used to compare NPE to Bayesian calibration are computed on another held-out validation dataset containing 50 samples. The validation dataset size is significantly different from the train/test dataset size because parameter inference with Bayesian calibration is several orders of magnitude slower than for NPE (see Sec.~\ref{sec:cost} for quantitative cost comparisons).  %job id 11759021

% A se with the caveat that $\theta$ is sampled so as to ensure that the total mass of Li (across electrodes) is the same between the charge and the discharge.  This is achieved by fixing all but one initial intercalation Li fraction value and adjusting the last one (the floating parameter) to enforce total Li mass conservation. If the floating parameter does not fall withing the bounds shown in Tab.~\ref{tab:priorcompds}, another one of the parameter is left floating and the procedure is restarted. The sequence of parameters adjusted is, in order, $[x_{\rm 0,c,ch}, x_{\rm 0,a,ch}, x_{\rm 0,a,dis}, x_{\rm 0,c,dis}]$, where $x_{\rm .,.,ch}$ (resp. $x_{\rm .,.,dis}$) denotes the charge (resp. the discharge) initial intercalation fractions. If no acceptable solution is found with any of the floating parameters, the data point is excluded from the dataset. This approach is a way to encode a physics constraints with the dataset which would be difficult to do without NPE. However, it is specific to the present case where charge and discharge are simulated separately. The dataset contains 93,000 pairs of $(\theta, x)$

\noindent \underline{XCEL dataset}: The physics-based model used here is the pseudo-2D model (P2D)~\cite{doyle1993modeling,fuller1994relaxation} during $\rm{C}/2$ discharge and a $4\rm{C}$ charge, which are used simultaneously ($n_{\rm obs} = 2$). The physics-based model is different than for the Comparison dataset to illustrate that NPE is agnostic to the physics-based model. This would not be the case if a physics-informed neural net~\cite{hassanaly2024pinn2} was used as a model surrogate. The same parameters as in the Comparison dataset are used, but $N_{\theta}=8$ to separate the charge and discharge initial lithium intercalation fractions in either electrode. Unless otherwise specified, the priors are given by Tab.~\ref{tab:priorcompds} (middle and right columns). 

The training dataset contains 90,000 pairs of $(\theta, x)$ and the test dataset contains 10,000 pairs of $(\theta, x)$. Generating both datasets required a 5~h-long run on 208 Intel Xeon Sapphire Rapids CPU cores. The cost increase as compared to the Comparison dataset is mainly due to the use of a P2D model as compared to the SPM, and the fact that one constant-current discharge and one constant-current charge are simulated for each $\theta$ (compared to only one constant-current discharge for the Comparison dataset). %job id 11759033 in case
The simulation dataset uses one of the cycling protocols of the XCEL experimental campaign, where reference performance tests (RPT) were intermittently conducted during cycle aging~\cite{weddle2023battery}. In particular, a slow discharge ($\rm{C}/20$) was used to extract loss of lithium inventory ($\rm{LLI}$) and loss of active material in the positive electrode ($\rm{LAM}_{\rm PE}$)~\cite{kim2022rapid}. 
Since the RPT test data is not used in the train/test datasets, the values of $\rm{LLI}$ and $\rm{LAM}_{\rm PE}$ are the validation metrics for this dataset.

\subsubsection{Convolutional neural posterior estimation (CNPE)}

The NPE must map an observation $x$ to the posterior $q_{\phi}(\theta|x)$. Throughout this manuscript, the parameter posterior is modeled as 
\begin{equation}
    \label{eq:post_approx}
    q_{\phi}(\theta|x) = \mathcal{N}\bigg(\mu(x),~\sigma(x) \operatorname{I_{N_{\theta}}}\bigg),
\end{equation}
where $\forall x, \mu(x), \sigma(x) \in  \mathbb{R}^{N_{\theta}}$ and $\operatorname{I}_{N_{\theta}}$ is the identity matrix, and $\mathcal{N}$ is a normal distribution. In other words, the posterior is an independent multivariate normal. By approximating $q_{\phi(\theta|x)}$ as an independent multivariate normal, only the mean $\mu(x)$ and standard deviation $\sigma(x)$ need to be determined. Other approximations are possible here, either by inferring more parameters of the posterior~\cite{villalobos2025neural} or by using probability transport methods~\cite{wildberger2023flow,papamakarios2021normalizing}. In Sec.~\ref{sec:comparison}, it is shown that Eq.~\ref{eq:post_approx} is a reasonable approximation for most of the inferred parameters in the present Li-ion battery example. 

\begin{figure}
    \centering
    \includegraphics[width=0.95\textwidth]{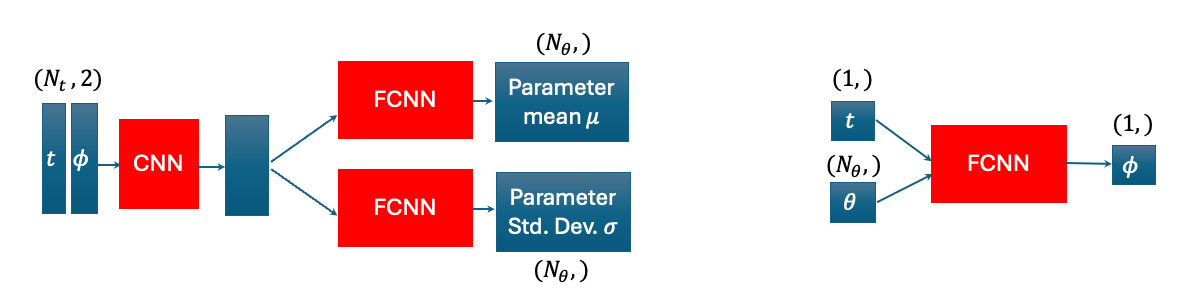}
    \caption{Schematic illustration of the neural net architectures used for CNPE (left) and the model surrogate (right). ``CNN" refers to convolutional layers, and ``FCNN" refers to fully connected layers.}
    \label{fig:arch}
\end{figure}

Approximating $\sigma(x)$ and $\mu(x)$ can be done by using a neural network that maps $x$ to $\mu(x)$ and $\sigma(x)$. Since $x$ is an ensemble of $n_{\rm obs}$ signals, choosing a convolutional neural network (CNN) for NPE allows for exploiting temporal correlation over the voltage signal to build $\mu(x)$ and $\sigma(x)$. The basic architecture of the network is schematically shown in Fig.~\ref{fig:arch} (left). In what follows, this network is referred to as \textit{Convolutional neural posterior estimation (CNPE)}.

\subsubsection{Bayesian calibration with surrogate likelihood}
\label{sec:surr_method}
Similar to Ref.~\cite{hassanaly2024pinn2}, the Bayesian calibration is conducted with a surrogate model in place of the physics-based model.

\begin{figure}
    \centering
    \includegraphics[width=0.4\textwidth]{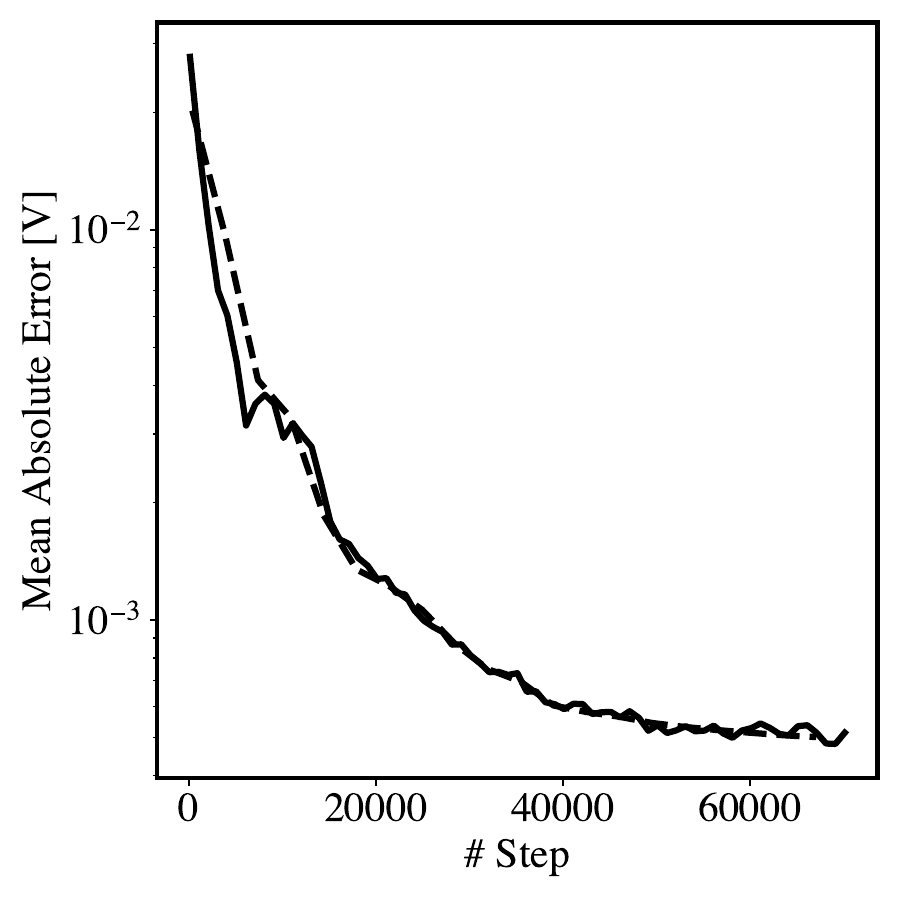}
    \includegraphics[width=0.4\textwidth]{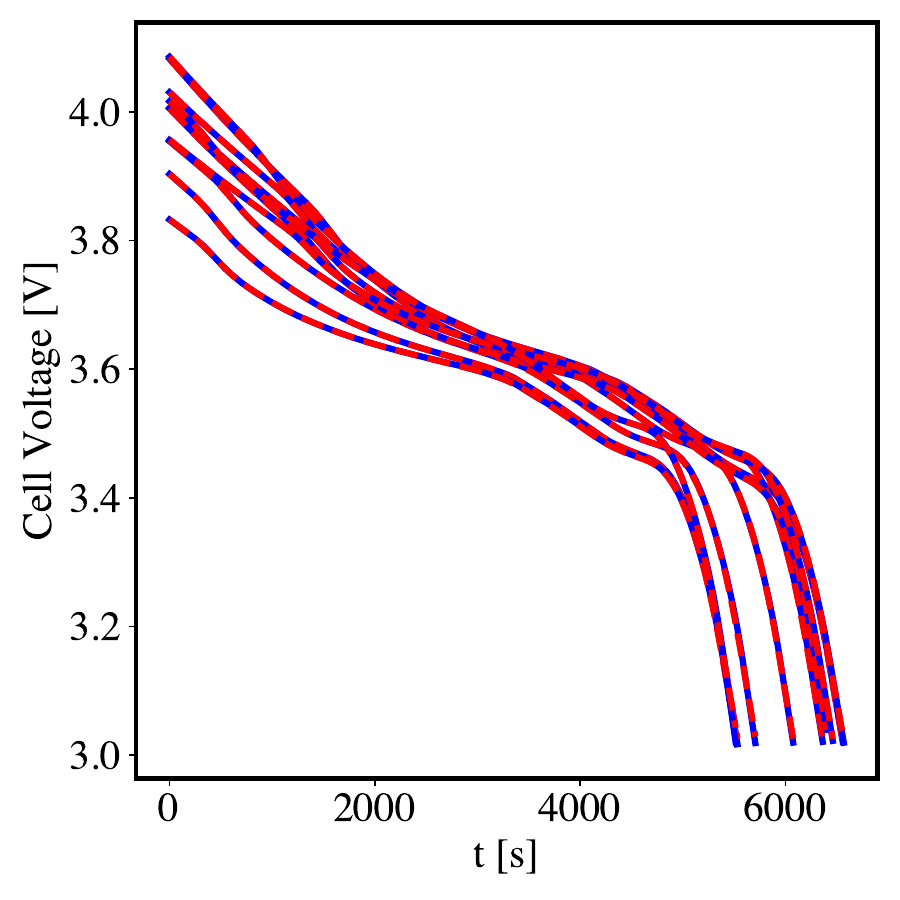}
    
    \caption{Training results of the discharge surrogate. Left: train (\mythickline{black}) and test (\mythickdashedline{black}) losses history. Right: discharge voltage curves obtained with different parameters $\theta$ with the SPM (\mythickline{blue}) and the surrogate (\mythickdashedline{red}).}
    \label{fig:surr_perf}
\end{figure}

\underline{Surrogate}: Two surrogates are trained: one that predicts the SPM constant current discharge at $\rm{C}/2$ and one  that predicts the SPM constant current charge at $4\rm{C}$. Compared to Ref.~\cite{hassanaly2024pinn1}, the surrogates are purely data-based and do not use a physics-informed loss. This has the advantage of simplifying the surrogate architecture as only voltage needs to be predicted by the surrogate. This simplification, however, comes at the expense of a data-intensive approach. Since Bayesian calibration is only used for low dimensional cases ($N_{\theta}=6$), a purely data-based approach is acceptable. The dataset used here is the Comparison dataset with several noteworthy differences:
\begin{itemize}[noitemsep,topsep=0em]
    \item Since the surrogate is meant to approximate the physics-based model, no noise model is used. The measurement noise is accounted for during the Bayesian calibration step (similar to \cite{hassanaly2024pinn2}).
    \item The surrogate ingests pointwise time and parameter values, meaning that each parameter set $\theta$ appears $N_t$ (here 256) times in the modified dataset.
\end{itemize}
The architecture of the surrogate is schematically shown in Fig.~\ref{fig:arch} (right), and the hyperparameters are given in ~\ref{app:hypersurr}. The same hyperparameters are used for the discharge and the charge surrogate. The network is a regression model trained with a mean absolute error loss. The train and test loss history for the discharge surrogate are shown in Fig.~\ref{fig:surr_perf}. The loss history suggests that the surrogate did not overfit. The surrogates reach a final mean absolute error (MAE) of 0.67 mV on the $\rm{C}/2$ discharge test dataset, and 0.39 mV on the $4\rm{C}$ charge dataset. In Fig.~\ref{fig:surr_perf} (right), ten random discharge prediction samples are shown. It can be seen that the temporal smoothness of the voltage curve is realistic and accurate despite training the surrogate pointwise. It can also be seen that the voltage prediction is sufficiently accurate to capture the effect of $\theta$ variations.
 
\underline{Bayesian calibration}: Bayesian calibration is done similarly to Ref.~\cite{hassanaly2024pinn2} using Hamiltonian Monte-Carlo \cite{hoffman2014no,numpyro}. The number of chains is set to 5, with 750 warmup samples and 100 posterior samples per chain. The voltage observations are subject to the same noise as the one used to train the CNPE. The likelihoods are computed with the surrogate described above. The likelihood uncertainty is calibrated as part of the Bayesian Calibration with a prior set to $\mathcal{U}(0.1~\rm{mV}, 3 ~\rm{mV})$. In Ref.~\cite{hassanaly2025bayesian}, it was shown that calibrating the likelihood noise is equivalent to treating it as a hyperparameter (like in Ref.~\cite{hassanaly2024pinn2}).

\section{Results}
\label{sec:results}

\subsection{Comparison between neural posterior estimation and Bayesian calibration}
\label{sec:comparison}

\begin{figure}
    \centering
    \includegraphics[width=0.4\textwidth]{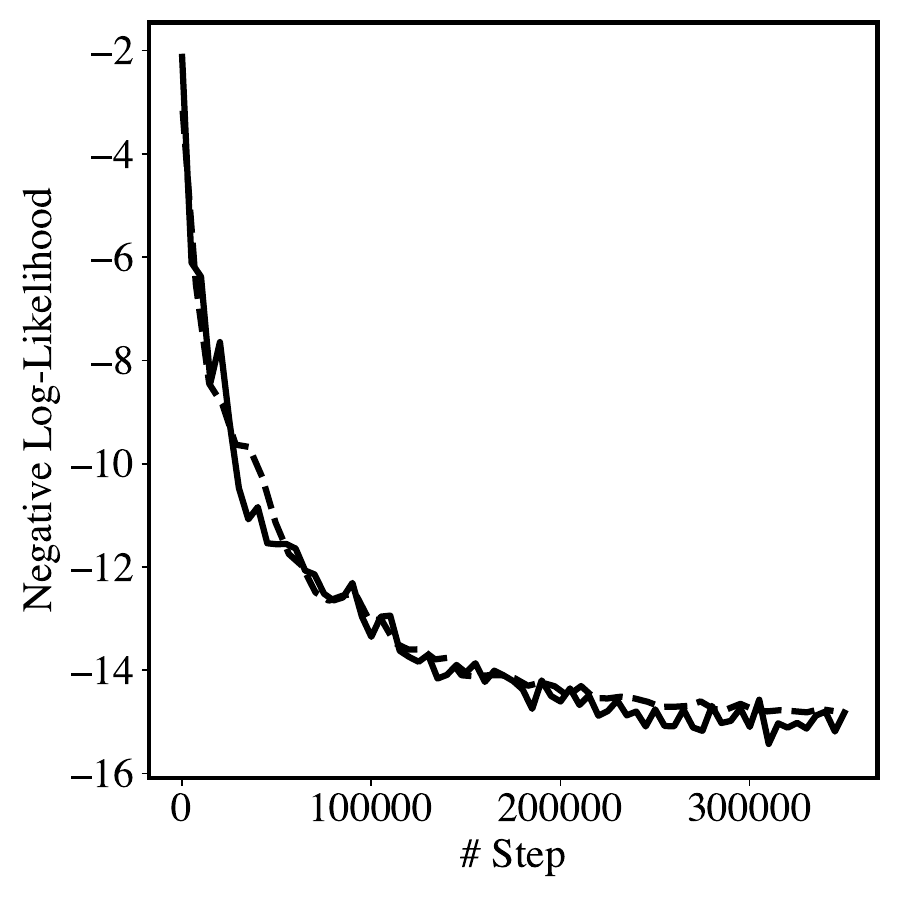}
    \caption{Train (\mythickline{black}) and test (\mythickdashedline{black}) losses history for CNPE trained on the discharge comparison dataset. NLL denotes the batch-averaged negative log-likelihood.}
    \label{fig:cnpe_perf}
\end{figure}

In this section, the parameter calibration results are presented for both the Bayesian calibration and the CNPE. The objective is to understand whether CNPE provides reasonable parameter distributions predictions even when using the variational approximation. The CNPE hyperparameters are obtained via hyperparameter tuning as described in \ref{app:hypercnpe}. Similar to the surrogate training, the loss function (Fig.~\ref{fig:cnpe_perf}) suggests that the training procedure converged and that the network did not overfit. 

\subsubsection{Accuracy}

This section presents the evaluation metrics used to compare the predictive abilities of CNPE and Bayesian calibration. All the metrics computed here are obtained on the 50 samples of the held-out validation discharge and charge datasets.

\noindent\paragraph{Parameter Error (PE)} The parameter error $\varepsilon_{\rm PE}$ metric evaluates the precision of the parameter distribution. It is computed by comparing the mean parameter predicted to the unknown true parameter values. The parameter error can be expressed as
\begin{equation}
    \varepsilon_{\rm PE} = \frac{1}{N_{\rm{test}}} \sum_{j=1}^{N_{\rm test}}  \frac{1}{N_{\theta}}  \sum_{i=1}^{N_{\theta}} \frac{1}{{A_i}}  |\theta_{j,i \rm true} - \mu_i(x)_j|   %\sum_{j=1}^{N_{\rm test}} | \theta_{j,i, \rm{true}} - \mu(x_i)_j|, 
\end{equation}
where $x$ is the observation, $A_i$ is the size of the support of the $i^{\rm th}$ parameter prior (the difference between the max and the min value of the $i^{\rm th}$ parameter), $\theta_{j,i \rm true}(x)$ is the ground truth for the parameters inferred, $\mu_i(x)_j$ is the mean $i^{\rm th}$ parameter value predicted for the $j^{\rm th}$ data point. The scaling with $A_i$ allows $\varepsilon_{\rm PE}$ to be expressed as a percentage error. A similar error can be defined based on the parameter samples instead of the mean parameter values
\begin{equation}
    \varepsilon_{\rm PE, s} = \frac{1}{N_{\rm{test}}} \sum_{j=1}^{N_{\rm test}}  \frac{1}{N_{\rm s}} \sum_{k=1}^{N_{\rm s}} \frac{1}{N_{\theta}}  \sum_{i=1}^{N_{\theta}} \frac{1}{{A_i}}  |\theta_{j,i \rm true} - \theta_i(x)_{j,k}|,   %\sum_{j=1}^{N_{\rm test}} | \theta_{j,i, \rm{true}} - \mu(x_i)_j|, 
\end{equation}
where $N_{\rm s}$ is the number of posterior samples, and $\theta_i(x)_{j,k}$ is the $k^{\rm th}$ sample of the posterior of $i^{\rm th}$ parameter given the $j^{\rm th}$ data point.

%The mean value of $\varepsilon_{\rm PE}$ over the held-out validation set of the Comparison dataset is reported in this section. 

\noindent\paragraph{Coverage Error (CE)} The coverage error $\varepsilon_{\rm CE}$ evaluates the accuracy of the predicted uncertainty. A conformal-prediction evaluation is used here, and can be seen as a frequentist metric for the statistics inferred \cite{angelopoulos2021gentle}. If the predicted standard deviations $\sigma(x)$ of the Gaussian posterior are accurate, then one standard deviation should cover true parameter values 68\% of the time, two standard deviations should cover true parameter values 95\% of the time, and three standard deviations should cover true values 99\% of the time. These target frequencies are called $f_{{\rm true},k}$, where $k$ is the number of standard deviations. We consider a frequency function $f_{i,k}$ that records how frequently (in percent), over the validation set, a parameter falls inside the range $\mu_i(x) \pm k\sigma_i(x)$. Then the coverage error metric is expressed as
\begin{equation}
    \varepsilon_{\rm CE} = \sum_{k=1}^3 |f_{\rm{true}, k} - \frac{1}{N_{\theta}} \sum_{i=1}^{N_{\theta}} f_{i,k} |.  
\end{equation}
Similar to the parameter error $\varepsilon_{\rm PE}$, the coverage error $\varepsilon_{\rm CE}$ can be expressed as a percent. In the case of the Bayesian calibration, the coverage error is expressed the same way, except percentiles are used instead of standard deviations.

\noindent\paragraph{Voltage Error (VE)} The voltage error $\varepsilon_{\rm VE}$ evaluates how good is the voltage fit that uses the predicted parameters. The voltage error can be expressed as
\begin{equation}
    \varepsilon_{\rm VE} = \frac{1}{N_{\rm s}}\sum_{s=1}^{N_{\rm s}} \frac{1}{N_{\rm t}} \sum_{l=1}^{N_{\rm t}}  |\mathcal{M}(\theta_s)_l - x_l|, 
\end{equation}
where $N_{\rm t}$ is the number of discretization steps for the voltage time series $x$, $N_{\rm s}$ is the number of samples drawn from the posterior, and $\mathcal{M}$ is the forward model (here, the SPM). Given the noise model, if model parameters are perfectly estimated, the voltage error is $\varepsilon_{\rm VE} = 0.72~\rm{mV}$, which can be considered the minimal error achievable. Here $N_{\rm s}=500$ since 5 chains with 100 samples each are drawn for the Bayesian calibration (Sec.~\ref{sec:surr_method}). The voltage error $\varepsilon_{\rm VE}$ metric is expensive to compute since it requires evaluating the physics-based model $\mathcal{M}$  for every parameter sample. To mitigate the computing cost associated with evaluating $\mathcal{M}$, the SPM is replaced with the surrogate trained in Sec.~\ref{sec:surr_method}.  The voltage error metric using the surrogate model is denoted as $\varepsilon_{\rm VE, Surr}$. The same metric using the physics-based model (denoted as  $\varepsilon_{\rm VE, Phy}$) is computed for comparison. When using the physics-based model, $N_{\rm s}$ is reduced to 50.  

\noindent\paragraph{Results} The accuracy metrics of both the Bayesian calibration and CNPE are shown in Tab.~\ref{tab:acc}, and the best-performing results are highlighted in bold. Overall, the results are consistent between the charge and the discharge Comparison datasets. Parameter errors $\varepsilon_{\rm PE}$ are small for both the CNPE and Bayesian calibration, but the CNPE consistently outperforms Bayesian calibration. A larger difference can be observed for coverage error $\varepsilon_{\rm CE}$, again with the CNPE outperforming the Bayesian calibration. In general, Bayesian calibration tends to produce uncertainty bands that are too tight around the parameters.
By contrast, voltage errors $\varepsilon_{\rm VE}$ are significantly smaller for the Bayesian calibration as compared to the CNPE. In particular, Bayesian calibration achieves $\varepsilon_{\rm VE, Surr}$ almost as small as the minimal achievable error ($0.72~{\rm mV}$). The higher CNPE voltage error $\varepsilon_{\rm VE}$ suggests that the CNPE generates conservative parameter estimates. Though the voltage errors from the physics-based model $\varepsilon_{\rm VE, Phys}$ are generally consistent with the surrogate $\varepsilon_{\rm VE, Surr}$ at charge conditions, a larger discrepancy is observed during discharge. This is especially apparent for the Bayesian calibration where actual voltage errors are 2$\times$ the ones predicted by the surrogate. This points towards an inaccuracy of the SPM surrogate at discharge conditions, and is discussed further in Sec.~\ref{sec:vi_approx}.

Overall, while the CNPE gives better parameter posterior estimates, it does not necessarily lead to the closest voltage match. This is a surprising result, suggesting that voltage matching alone is not sufficient to decide about an accurate model parameterization. There are likely two reasons behind this finding. First, Bayesian calibration discriminates parameter samples based on the predicted voltage, while CNPE is trained to only predict the parameter values. In particular, if parameter errors can compensate for experimental noise, MCMC can overfit the noise with the parameters. This effect is demonstrated in~\ref{app:noise} where the synthetic noise levels are varied. Second, voltage errors $\varepsilon_{\rm VE}$ grow rapidly as the parameters vary. Since CNPE approximates possibly complex distributions with a Gaussian, the posterior can include parameter values that lead to large voltage errors. This is supported by the fact that while mean and median voltage errors are similar for the Bayesian calibration (not shown in Tab.~\ref{tab:acc} for the sake of concision), median errors are significantly smaller in the case of CNPE. 

\begin{table}[h!]
\centering
 \begin{tabular}{ |c|c|c|c|c| } 
        \hline
        Metric &  \multicolumn{2}{|c|}{Bayesian calibration} & \multicolumn{2}{|c|}{CNPE} \\
        \hline
         & Discharge $\rm{C}/2$ & Charge $4\rm{C}$ & Discharge $\rm{C}/2$ & Charge $4\rm{C}$  \\
        \hline
         Parameter error (mean) $\varepsilon_{\rm PE}$   &  $8.1\%$ & $6.14\%$  & \boldsymbol{$7.68\%$} & \boldsymbol{$4.41 \%$}\\
         Parameter error (sample) $\varepsilon_{\rm PE, s}$   &  \boldsymbol{$9.98\%$} & $7.33\%$  & $10.58\%$ & \boldsymbol{$6.32 \%$}\\ 
         Coverage error $\varepsilon_{\rm CE}$   &  $62.45\%$  &  $70.45\%$ & \boldsymbol{$9.91\%$} &  \boldsymbol{$1.68\%$} \\ 
         Voltage error (surrogate) $\varepsilon_{\rm VE, Surr}$   &   \boldsymbol{$0.801~{\rm{mV}}$} & \boldsymbol{$0.792~{\rm{mV}}$}  & $3.096$~mV ($2.328$ mV)   & $5.493$ mV ($2.862$ mV)  \\ 
         Voltage error (physics-model) $\varepsilon_{\rm VE, Phy}$   &   \boldsymbol{$1.751 {\rm{mV}}$} & \boldsymbol{$ 0.919~{\rm{mV}}$}  & $4.624$~mV ($3.707$~mV)   & $5.773$~mV ($2.598$~mV)  \\ 
       
        \hline

        \end{tabular}
\caption{Accuracy metric comparisons between CNPE and Bayesian calibration on the validation set of the Comparison dataset. Best results are highlighted in bold. For CNPE, median voltage errors are reported in parentheses.}
\label{tab:acc}
\end{table}

%\todo{Rewrite based on the results once runs finish}
%uses the voltage fits in the likelihood function. Since the mapping between parameter and voltage is strongly non-linear, a small error in the posterior support (expected for CNPE), can have large effects on the voltage prediction. This result also suggests that it is inaccurate to equate parameter fits and voltage fit \cite{...}. 

\subsubsection{Suitability of the variational approximation}
\label{sec:vi_approx}

\begin{figure}
    \centering
     \includegraphics[width=0.49\textwidth]{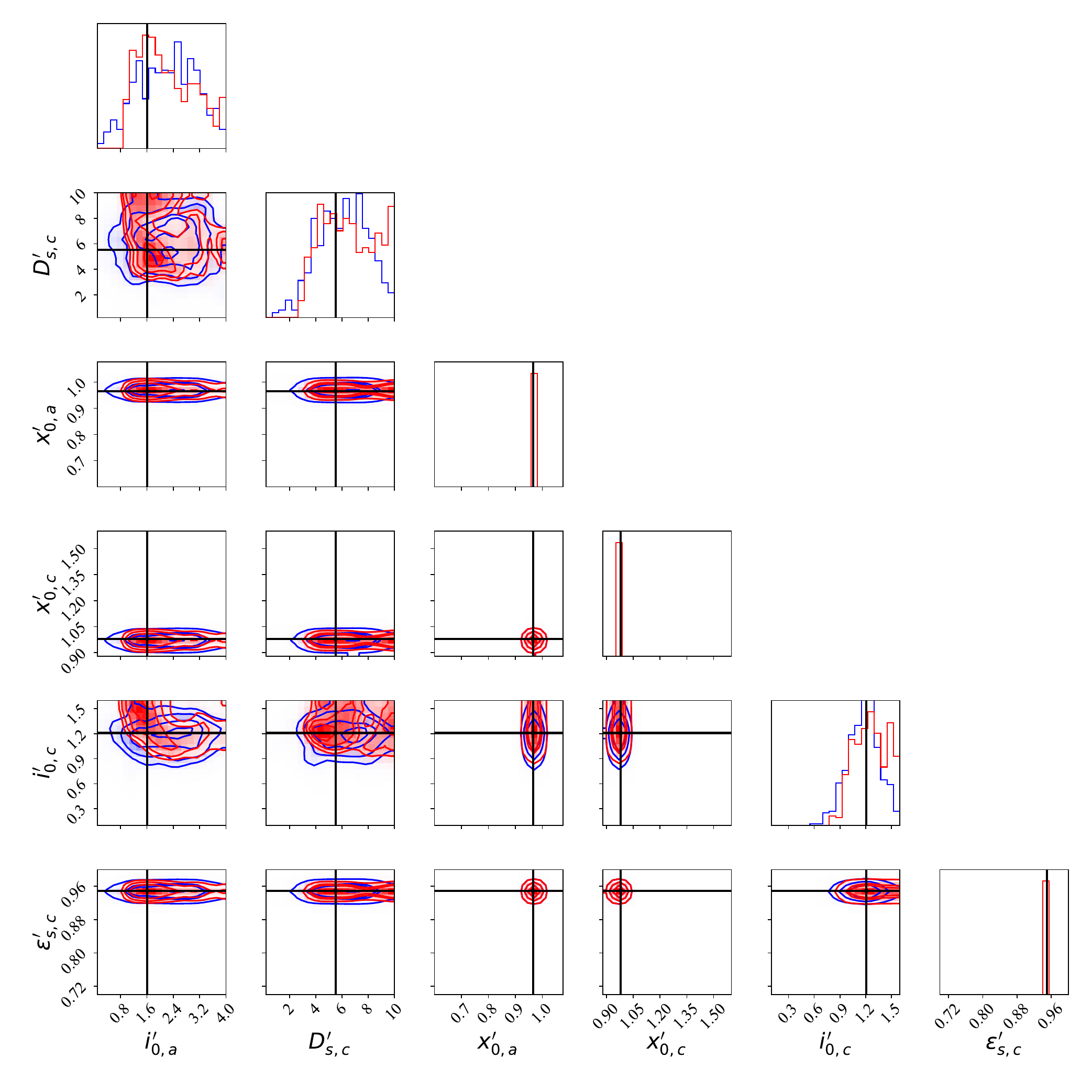}
     %overlapped_corner_discharge_new_prune
     \includegraphics[width=0.49\textwidth]{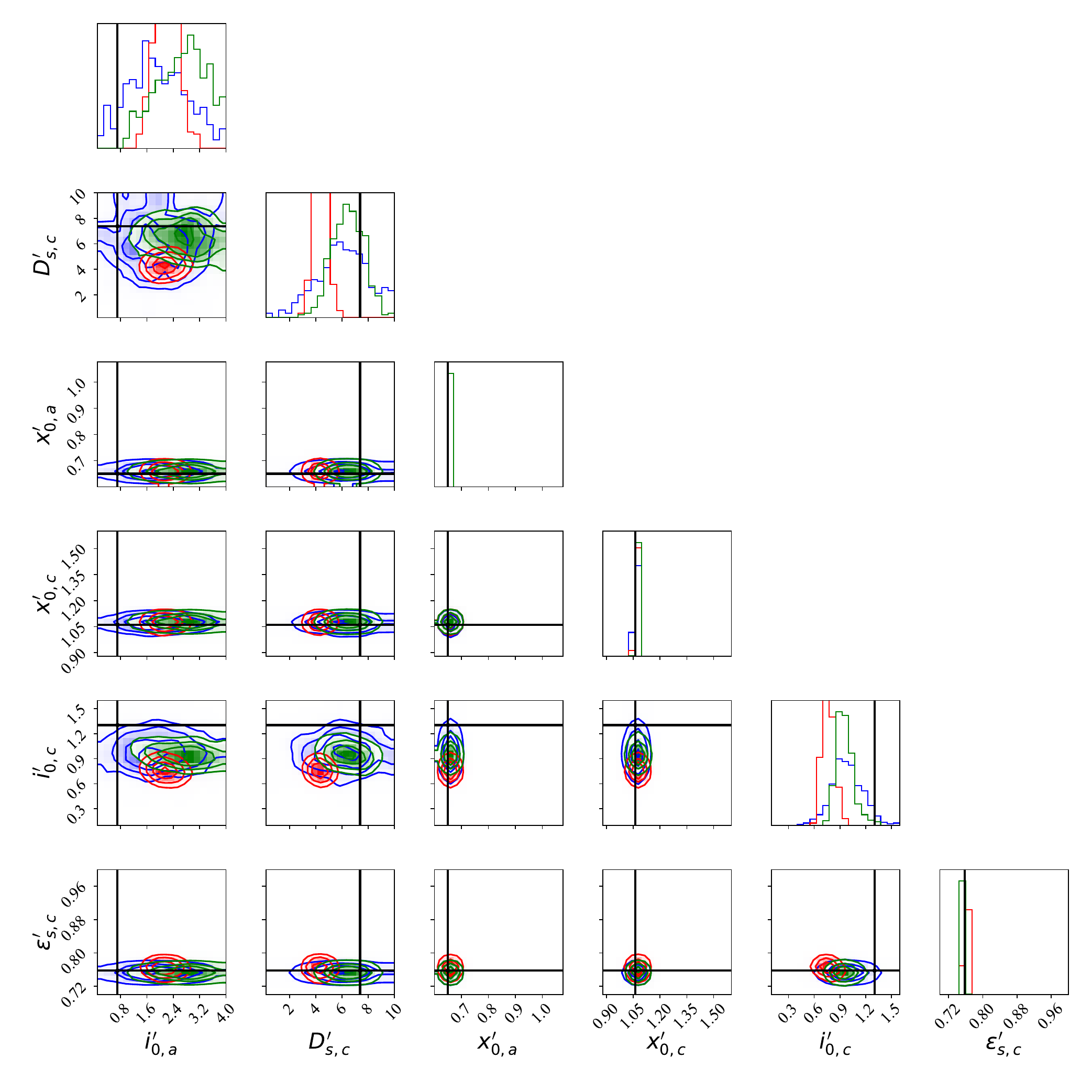}

     %overlapped_corner_discharge_new_prune_57

    \caption{Left: example of near-matching posteriors between Bayesian calibration (\mythickline{red}) and CNPE (\mythickline{blue}). Right: example of posterior mismatch between Bayesian calibration (\mythickline{red}) and CNPE (\mythickline{blue}). Bayesian calibration done with a more accurate surrogate is also shown (\mythickline{green}). Solid lines (\mythickline{black}) denote the true values.}
    \label{fig:corner}
\end{figure}

The CNPE can be seen as an approximation of Bayesian calibration that relies on a predetermined functional form for the posterior. In the present section, the objective is to understand whether the variational approximation is appropriate. Figure~\ref{fig:corner} shows the posterior predicted with Bayesian calibration and CNPE for two data points of the discharge Comparison dataset. Though MCMC can sample complex distributions, the posterior predicted by MCMC is visually close to the one predicted by CNPE. In particular, the joint parameter PDFs predicted by MCMC are almost all ellipsoids which suggest that the variational approximation is appropriate. For joint PDFs that involve the efficiency factor of the anode exchange current density $i_{\rm 0, a}'$ and the efficiency factor of the solid-phase cathode diffusion coefficient $D_{\rm s, c}'$, the independent Gaussian approximation is less appropriate, and the posterior of CNPE includes samples rejected by MCMC. 
This effect could lead to higher voltage errors for the CNPE approach. Importantly, the variational approximation is appropriate for the efficiency factors of the starting anode intercalation fraction $x_{\rm 0, a}'$, the cathode starting intercalation fraction $x_{\rm 0, c}'$ and the cathode's solid-phase volume fraction $\varepsilon_{\rm s, c}'$, which can be used to compute loss of lithium inventory (LLI) and loss of active material (LAM) (see Sec.~\ref{sec:application_exp} for the expression).

The right plot shows another example where a particular voltage trace results in a parameter posterior for which CNPE and MCMC differ more significantly. However, the variational approximation is still appropriate in general (the MCMC joint posteriors are still ellipsoids). 
The mean parameters predicted by MCMC for the efficiency factor of the anode exchange current density $i_{\rm 0, a}'$, the cathode exchange current density $i_{\rm 0, c}'$ and the cathode solid-phase diffusion coefficient $D_{\rm s, c}'$ is different than the one predicted by CNPE and the true values. This points to the fact that the surrogate model may be inaccurate at the parameter values considered. Since the optimal hyperparameters for the surrogate hit the boundary of the search space (\ref{app:hypersurr}), another surrogate was retrained with a maximal number of neurons per hidden layer of 4096, and a batch size of $2^{14}$. This new surrogate led to a test error of $0.39$~mV on the test discharge comparison dataset. The posterior obtained with that new surrogate is shown in Fig.~\ref{fig:corner} (right) and is closer to the CNPE. However, these results indicate that MCMC can be strongly affected by even small errors in the surrogate model.   

A quantitative comparison of the distributions can be obtained using an earth-moving Wasserstein distance based on posterior samples~\cite{flamary2021pot}, which can be used to compare how far PDFs are from one another (i.e., measure a distance between PDFs). Intuitively, the Wasserstein distance computes the minimal amount of work needed to transform one pile of dirt (a PDF) into another pile of dirt (another PDF)~\cite{bonneel2011displacement}. To make the units of the Wasserstein distance interpretable, we report 
\begin{equation}
    \varepsilon_{\rm W} = \frac{\sum_{i=0}^{N_{\rm val}} W(P_{\rm MCMC,i},P_{\rm CNPE, i}) } {\sum_{i=0}^{N_{\rm val}}  W(P_{\rm MCMC,i},P_{\rm prior, i})},
\end{equation}
where $W(P,Q)$ denotes the Wasserstein distance between $P$ and $Q$, $P_{\rm MCMC}$ is the posterior obtained via Bayesian calibration, $P_{\rm CNPE}$ is the posterior obtained with CNPE, and  $P_{\rm prior}$ is the prior. The values of $\varepsilon_{\rm W}$ are shown in Tab.~\ref{tab:wass}. For the joint 6-dimensional PDF, about $30\%$ error is observed for both the charge and the discharge cases. However, when looking at the marginal distributions, it is clear that the errors vary significantly across parameters. In particular, the variational approximation mostly affects the efficiency factor of the anode exchange current density $i_{\rm 0, a}'$ and the cathode solid-phase diffusion coefficient $D_{\rm s, c}'$, which is what was observed visually in Fig.~\ref{fig:corner}. During discharge, the high values of $\varepsilon_{\rm W}$ for the efficiency factor of the anode exchange current density $i_{\rm 0, a}'$ and of the cathode solid-phase diffusion coefficient $D_{\rm s, c}'$ are mainly due to the fact that the posteriors predicted with MCMC are close to the priors (i.e., constant current discharge is almost uninformative about $i_{\rm 0, a}'$ and $D_{\rm s, c}'$). For the parameters involved in LLI and LAM calculations, errors are quantitatively the smallest, which also echoes the results presented visually in Fig.~\ref{fig:corner}.

\begin{table}[h!]
\centering
 \begin{tabular}{ |c|c|c| } 
        \hline
        Parameter &  Discharge $\rm{C}/2$  & Charge $4\rm{C}$ \\
        \hline
         Joint & $34.20\%$ & $27.50\%$    \\
        \hline
        $i_{0,\rm{a}}'$ & $101.33\%$ &  $62.14\%$  \\
        $D_{\rm s,c}'$ & $109.01\%$ &  $57.06\%$ \\
        \boldsymbol{$x_{0,\rm{a}}'$} & \boldsymbol{$1.72\%$} & \boldsymbol{$2.77\%$} \\
        \boldsymbol{$x_{0,\rm{c}}'$} & \boldsymbol{$1.21\%$} & \boldsymbol{$4.49\%$} \\
        $i_{0,\rm{c}}'$ & $32.18\%$ & $24.23\%$ \\
        \boldsymbol{$\varepsilon_{\rm{s, c, AM}}'$} & \boldsymbol{$3.64\%$} & \boldsymbol{$5.89\%$} \\
        \hline

        \end{tabular}
\caption{Wasserstein distance $\varepsilon_{\rm W}$ computed for the joint and the marginal posteriors. Parameters involved in the LLI and LAM calculation are indicated in bold.}
\label{tab:wass}
\end{table}

\subsubsection{Cost}
\label{sec:cost}

In the case of CNPE, the main costs are related to the data generation. Generating realizations of the SPM model with constant-current cycling required 52 CPUh. Using a P2D model would have increased the data generation cost 10$\times$. Using cycling protocols typical of EVs would have further increased the data generation cost by two orders of magnitude. However, data generation is embarrassingly parallel, which makes the wall-clock time acceptable. By comparison, though the Bayesian calibration approach requires data for the surrogate training, the amount of data needed can be made significantly smaller with physics-informed regularization~\cite{hassanaly2024pinn1,hassanaly2024pinn2,brendel2025parametrized}. The CNPE method is generally faster to train because it preserves the temporal structure of the voltage data. Here, it took 4~GPUh to train a CNPE. By comparison, a surrogate for Bayesian calibration took 24~GPUh.

The key benefit of CNPE is for inference costs, also noted by Zhang et al.~\cite{zhang2026discovery}. The posterior predicted by the CNPE can be generated with one forward pass, taking about $100~\rm{ms}$ on a GPU. By comparison, constructing the posterior with Bayesian calibration requires MCMC, taking on average $15~\rm{min}$ per chain.  

\subsubsection{Interpretability}

Since CNPE directly maps voltage observations $N_x$ to the mean parameter values $\mu(x)$ and uncertainties $\sigma(x)$, the CNPE method offers advantages for model interpretability, which are described and demonstrated hereafter. 

\begin{figure}
    \centering
    \includegraphics[width=0.8\textwidth]{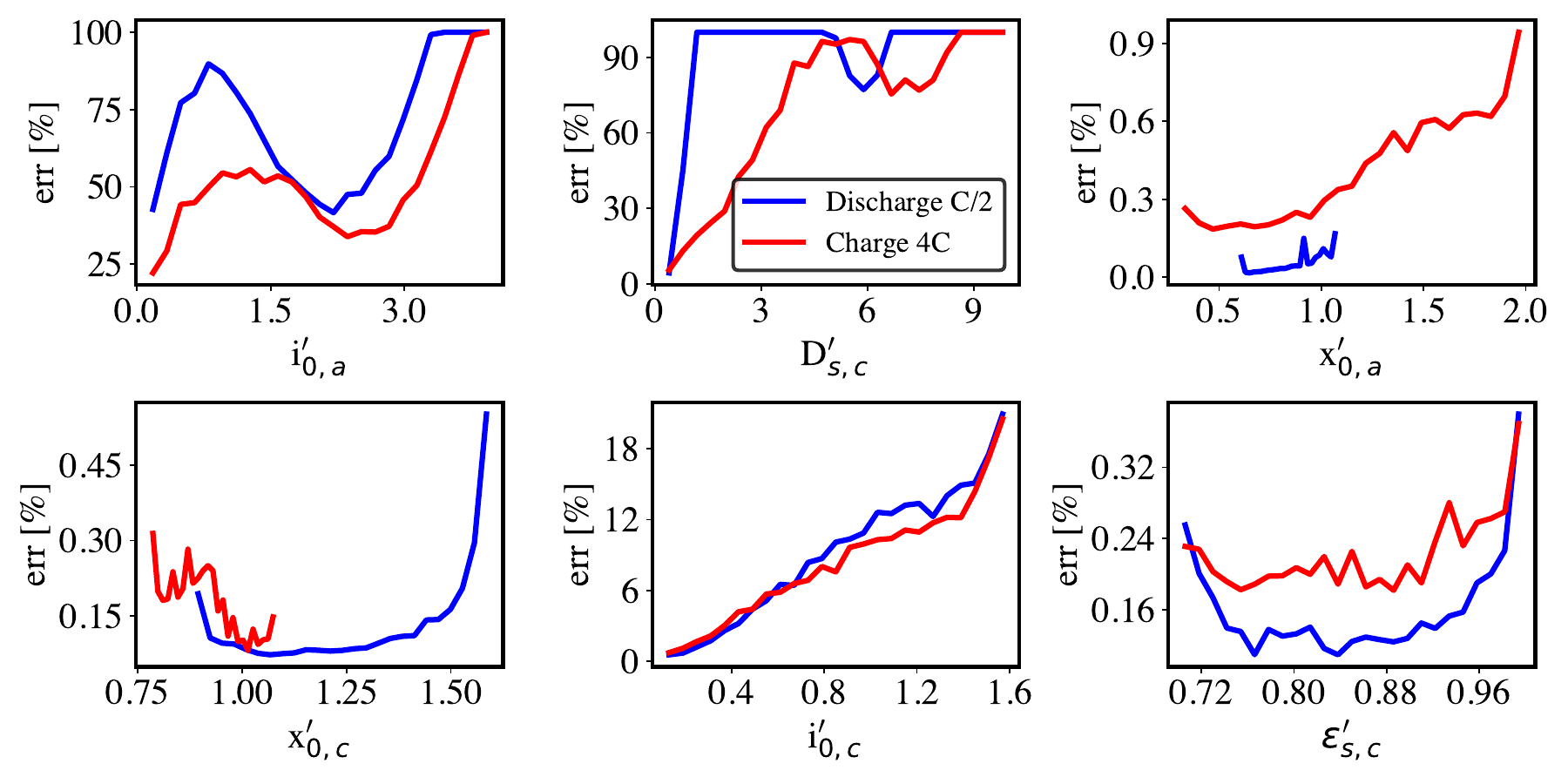}
    \caption{Conditional average of the relative error for each calibrated parameter evaluated on the discharge dataset (\mythickline{blue}) and the charge dataset (\mythickline{red}).}
    \label{fig:error_par_space}
\end{figure}

\noindent\paragraph{Granular model interrogation} Since the inference costs of CNPE are so low compared to Bayesian calibration, it becomes easy to extensively test where CNPE is accurate in parameter space. This is difficult to achieve with Bayesian calibration since each datapoint $P_{\rm MCMC}$ requires minutes of inference time. Hereafter, the CNPEs trained on the comparison dataset for $\rm{C}/2$ discharge and $4\rm{C}$ charge are evaluated on the test datasets (10,000 points). Since the test set is so large, one can compute conditional averages of the error across parameter space, and identify where errors are expected to be large. Figure~\ref{fig:error_par_space} shows the conditional parameter relative errors ($\rm{err}$). 

As both efficiency factors for anode and cathode exchange current density ($i_{\rm 0,a}'$ and $i_{\rm 0,c}'$, respectively) increase, they become harder to identify. The $4\rm{C}$ charge conditions are also clearly more informative about exchange current density than the $\rm{C}/2$ discharge conditions (i.e., the blue bars are higher than the red). The exchange current densities could likely be estimated with lower error if the data being calibrated varied the current density over time as seen in Li et al.~\cite{li2022data}. Errors incurred when estimating the efficiency factor of the cathode solid-phase diffusivity $D_{\rm s,c}'$ are also higher for discharge than for charge, especially at low diffusivities, likely due to the higher C-rate (current density) used for the charge. Conversely, the lower C-rate discharge conditions are more informative about efficiency factors for both starting intercalation fractions $x_0'$ and active material volume fraction $\varepsilon_{\rm s}'$, across the parameter space.   

\begin{figure}
    \centering
    \includegraphics[width=0.4\textwidth]{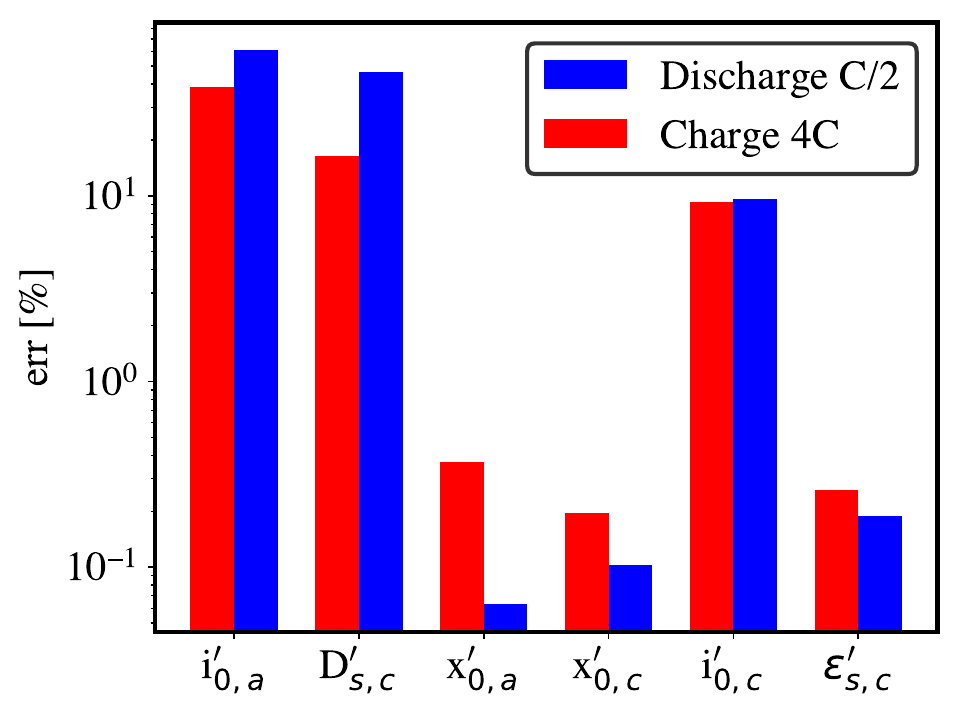}
    \includegraphics[width=0.4\textwidth]{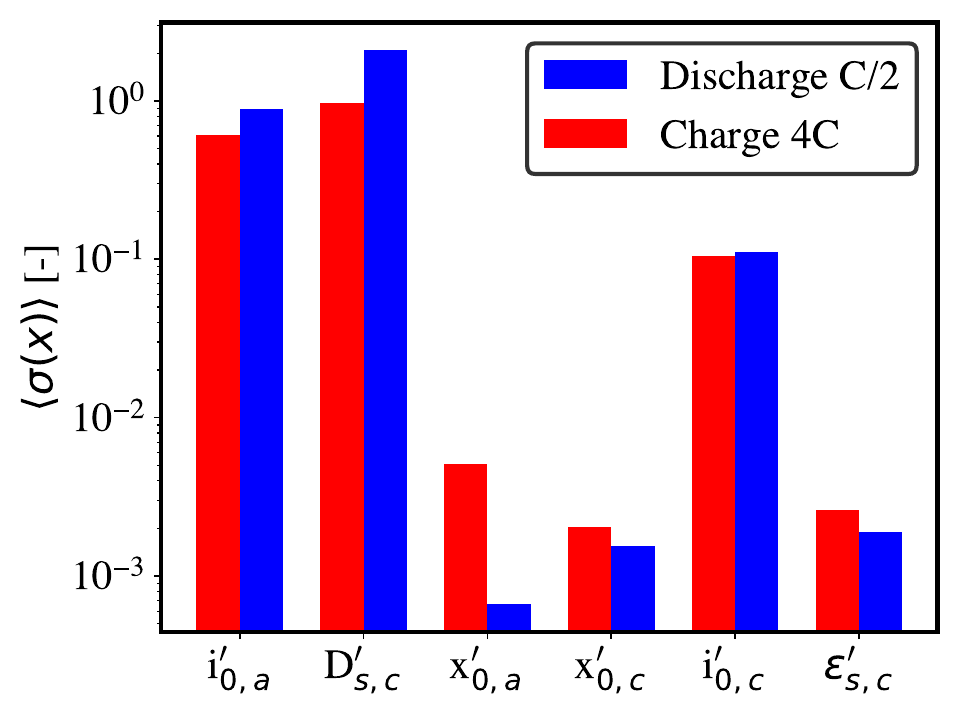}
    \caption{For the Comparison charge and discharge test datasets. Left: average relative errors for each parameter inferred. Right: average predicted parameter standard deviation.}
    \label{fig:ave_id}
\end{figure}

Figure~\ref{fig:ave_id} (left) illustrates the averaged conditional parameter relative errors (${\rm err}$) over the test datasets.  By inspecting this average over the test datasets, one can also identify which parameters are easy or hard to identify with different cycling protocols. A key benefit of CNPE is that it also outputs a predicted parameter standard deviation $\sigma(x)$, without needing to access the true parameter values. Specifically, one can also average $\sigma(x)$ over the test set for each parameter space (denoted as $\langle \sigma(x) \rangle$) and compare it to the actual average relative errors (cf., Fig.~\ref{fig:ave_id}(right)). Comparing Fig.~\ref{fig:ave_id} left and right, it is clear that the predicted variance indicates how much error is actually incurred for each parameter.

\noindent\paragraph{SHAP analysis} Unlike Bayesian calibration, CNPE constructs a direct mapping from voltage observations $N_x$ to parameter mean values $\mu(x)$. This mapping allows for using several well-established model explainability tools~\cite{perr2025applications}. Specifically, the \textit{SHAP analysis}~\cite{lundberg2017unified} can highlight which parts of the voltage curves inform specific parameters.   

\begin{figure}
    \centering
    \includegraphics[width=0.4\textwidth]{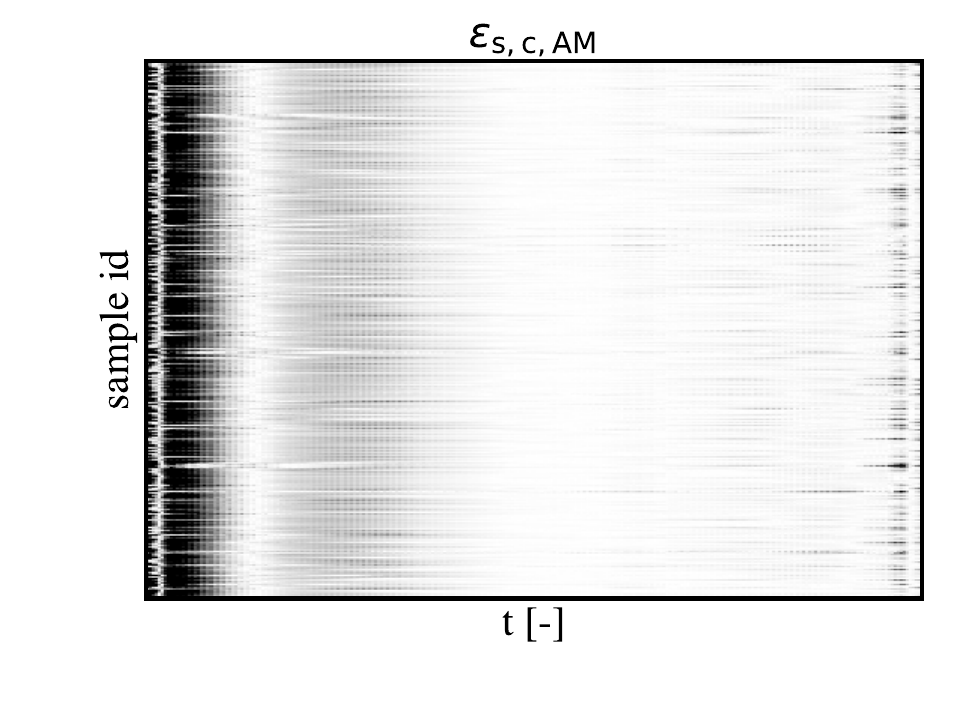}
    \includegraphics[width=0.4\textwidth]{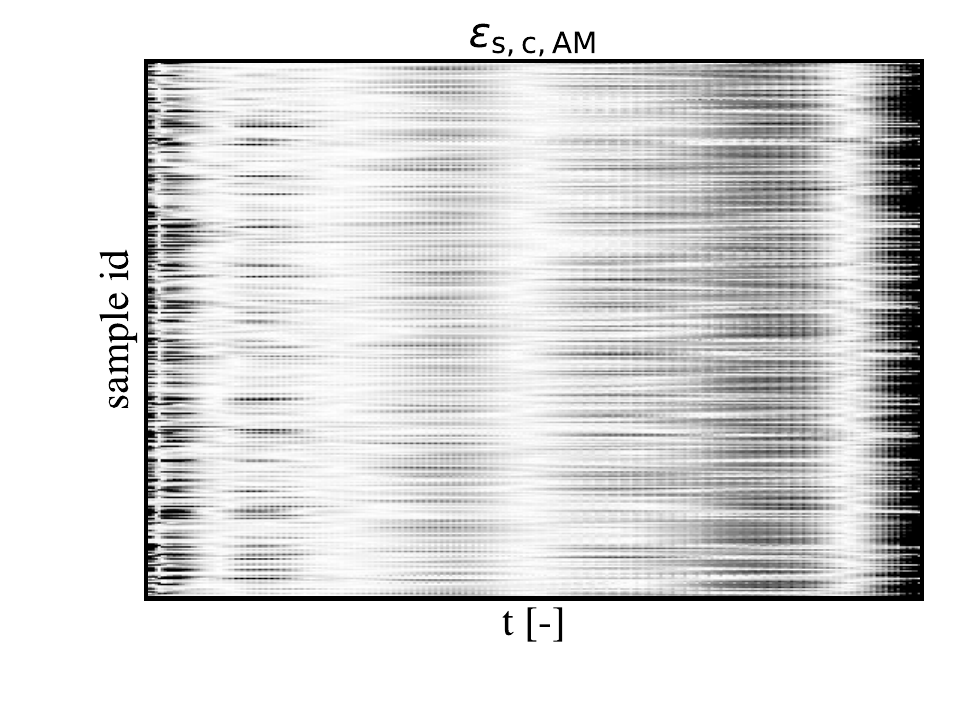}
    \caption{Absolute SHAP value obtained for 1000 discharge (left) and charge (right) realizations of the Comparison dataset, where the target is the mean value of $\varepsilon_{\rm s,c, AM}$. The darker, the higher the absolute SHAP value, i.e the more informative for $\varepsilon_{\rm s,c, AM}$.}
    \label{fig:shap}
\end{figure}

The absolute SHAP values  associated with the prediction of the mean value of the cathode solid-phase active material volume fraction $\varepsilon_{\rm s,c, AM}$ are shown for 1000 discharge and charge voltage curves. The background data contains 20,000 training curves (limited by the GPU memory). Figure~\ref{fig:shap} shows the results for all 1000 curves by stacking them over the ``sample id" axis and rescaling the time axis with respect to the end-of-charge and end-of-discharge time. Locations of high absolute SHAP values indicate parts of the voltage curve that are influential for the target value (here the mean value of $\varepsilon_{\rm s,c, AM}$). In general, the beginning of discharge and the end of charge, when the cathode is mostly delithiated, are mainly used by the CNPE network to infer the cathode active-material volume fraction $\varepsilon_{\rm s,c, AM}$. 

At minimum, this analysis can help ensure that the parameter predictions are grounded in physical considerations. At best, this analysis can be useful for feature identification~\cite{shrikumar2017learning}, which could help design informative cycling protocols~\cite{brendel2025parametrized} that highlight internal Li-ion battery parameters.

\subsection{Application to experimental data}
\label{sec:application_exp}

\subsubsection{LLI and LAM predictions}
\begin{figure}
    \centering
    \includegraphics[width=0.45\textwidth]{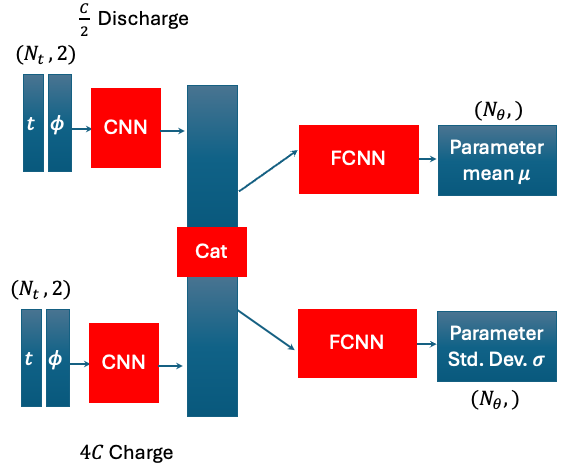}
    \caption{Schematic illustration of the modified architecture for CNPE. ``CNN'' refers to convolutional layers, ``FCNN'' refers to fully connected layers, and ``Cat'' refers to a concatenation operation.}
    \label{fig:arch_cat}
\end{figure}

In the present section, the CNPE method is demonstrated on the experimentally measured XCEL dataset. Here, the objective is to predict loss of lithium inventory ($\rm{LLI}$) and loss of active material in the positive electrode ($\rm{LAM}_{\rm PE}$) from the parameters inferred. At cycles $\{1, 26, 76, 126, 226, 326, 376, 451, 526\}$, the voltage data from the aging cycle (including constant-current charge and discharge) are extracted and passed through a CNPE. The architecture of the CNPE is modified as shown in Fig.~\ref{fig:arch_cat} so that parameters are predicted based on two voltage traces. Specifically, the charge and discharge events are passed through two separate convolutional layers and the results are concatenated before predicting the parameter posteriors.  

From the inferred parameters shown in Tab.~\ref{tab:priorcompds}, $\rm{LLI}$ and $\rm{LAM}_{\rm PE}$ can be reconstructed as 
\begin{equation}
    \rm{LAM_{PE}} = 1-\frac{\overline{\varepsilon_{\rm s,c,AM}}}{\overline{\varepsilon_{\rm s,c,AM}(\rm{BOL})}},
\end{equation}
and 
\begin{equation}
    \label{eq:lli}
    {\rm{LLI}} = 1-\frac{\overline{x_{0,\rm{a}}} C_{\rm s,a,max} L_{\rm{a}} \varepsilon_{\rm s,a, AM}
        + \overline{x_{0,\rm{c}}} C_{\rm s,c,max} L_{\rm{c}} \overline{\varepsilon_{\rm s,c, AM}}}{\overline{x_{0,\rm{a}}(\rm{BOL})} C_{\rm s,a,max} L_{\rm{a}} \varepsilon_{\rm s,a, AM}
        + \overline{x_{0,\rm{c}}(\rm{BOL})} C_{\rm s,c,max} L_{\rm{c}} \overline{\varepsilon_{\rm s,c, AM}(\rm{BOL})}},
\end{equation}
where $L_{\rm{a}}$ is the anode thickness, $L_{\rm{c}}$ is the cathode thickness, $C_{\rm s,a,max}$ and $C_{\rm s,c,max}$ are the maximum solid-phase lithium concentrations in the anode and cathode, $\varepsilon_{\rm s,a, AM}$ is the anode active-material volume fraction, $\rm{BOL}$ denotes beginning-of-life values, and $\overline{(.)}$ denotes parameters that are scaled with the coefficients inferred. The LLI definition is subject to ambiguity since it can be defined from the initial intercalation fractions obtained at charge or discharge. Ideally, both definitions of LLI (denoted as $\rm{LLI}_{\rm ch}$ for charge and $\rm{LLI}_{\rm dis}$ for discharge) should match, and are shown in the plots hereafter. 

\begin{figure}
    \centering \includegraphics[width=0.49\textwidth]{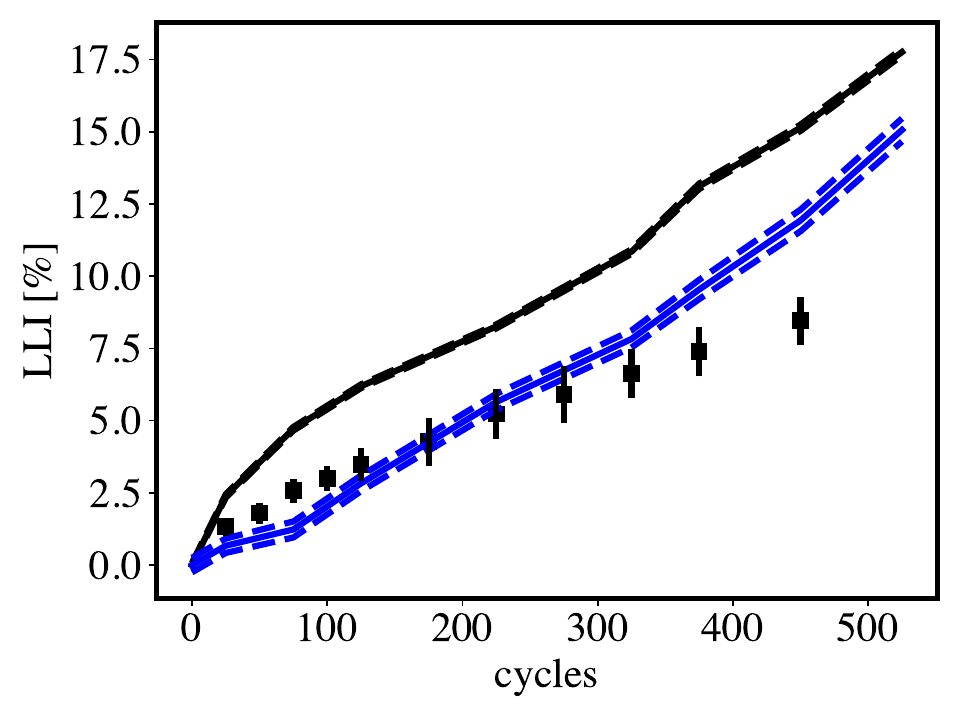}
    \includegraphics[width=0.49\textwidth]{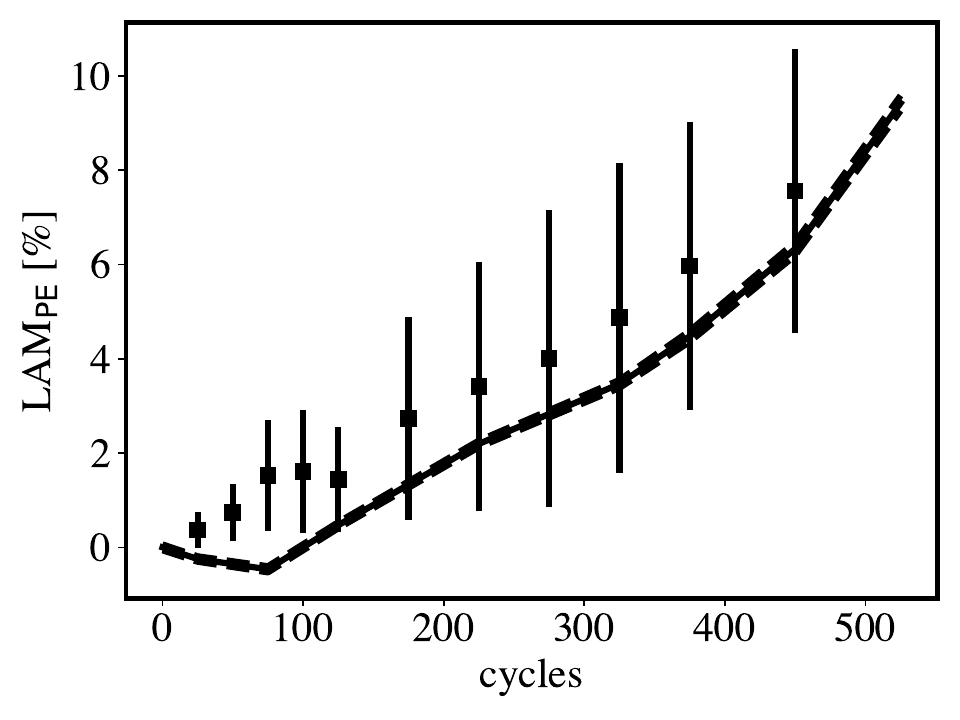}
    \caption{LLI (left) and $\rm{LAM}_{\rm PE}$ (right) measured during cycle aging (\mythickbarredsquare{black}{black}) and predicted mean (\mythickline{black}) and standard deviation (\mythickdashedline{black}). For the loss of lithium inventory (left), $\rm{LLI}_{\rm ch}$ (\mythickline{blue}) and $\rm{LLI}_{\rm dis}$ (\mythickline{black}) are shown.}
    \label{fig:LLILAM}
\end{figure}

The prediction of LLI and $\rm{LAM}_{\rm PE}$ are shown in Fig.~\ref{fig:LLILAM}. Although each cycle is treated independently from the other, LLI and $\rm{LAM}_{\rm PE}$ vary almost always monotonically, as expected from the experiment. Overall, the parameters allow to infer the rate of change of LLI and $\rm{LAM}_{\rm PE}$ and their magnitude. As the cell ages, the prediction of LLI departs more and more from the experimental data, which might indicate that a phenomenon not accounted for in the P2D model is occuring. In addition, for LLI, there is a clear difference that can be observed between $\rm{LLI}_{\rm ch}$ and $\rm{LLI}_{\rm dis}$. The focus of the rest of this section is to resolve this discrepancy.

\subsubsection{Lithium inventory conservation constraint}

\begin{figure}
    \centering \includegraphics[width=0.49\textwidth]{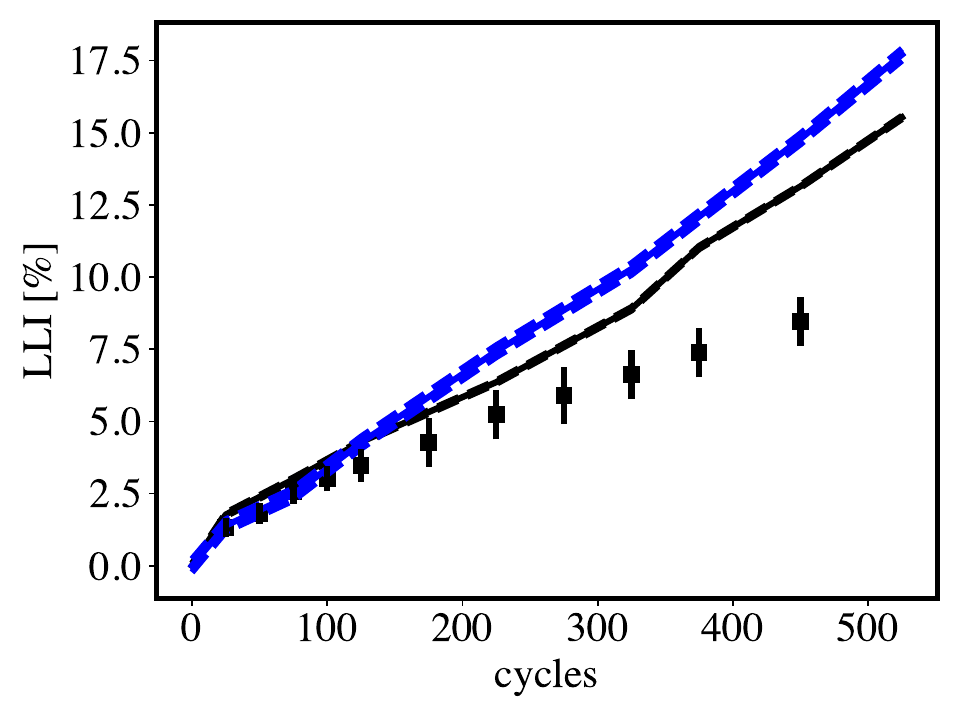}
    \includegraphics[width=0.49\textwidth]{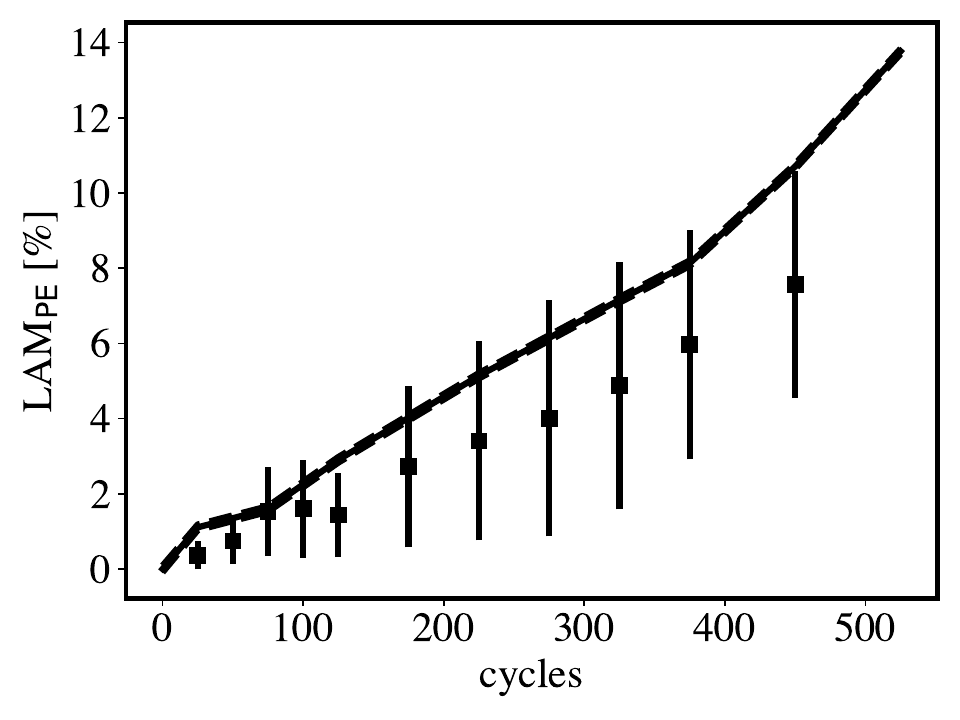}
    \caption{LLI (left) and $\rm{LAM}_{\rm PE}$ (right) measured during cycle aging (\mythickbarredsquare{black}{black}) and predicted mean (\mythickline{black}) and standard deviation (\mythickdashedline{black}), with Li conservation enforced through the prior. For the loss of lithium inventory (left), $\rm{LLI}_{\rm ch}$ (\mythickline{blue}) and $\rm{LLI}_{\rm dis}$ (\mythickline{black}) are shown.}
    \label{fig:LLILAM_licons}
\end{figure}

We propose to include a Li conservation constraint through the parameter priors. Compared to Bayesian calibration, the prior is represented by the dataset itself. We construct a new training dataset where the parameters $\theta$ are sampled so as to ensure that the total Li inventory is the same between the charge and the discharge (i.e., enforce LLI$_{\rm ch}$ = LLI$_{\rm dis}$ for the same cycle).  This is achieved by fixing all but one initial intercalation Li fraction value and adjusting the last one (the floating parameter) to enforce total Li mass conservation. If the floating parameter does not fall within the bounds shown in Tab.~\ref{tab:priorcompds}, another one of the parameters is left floating and the procedure is restarted. The sequence of parameters adjusted is, in order, $[x_{\rm 0,c,ch}', x_{\rm 0,a,ch}', x_{\rm 0,a,dis}', x_{\rm 0,c,dis}']$, where $x_{\rm .,.,ch}'$ (resp., $x_{\rm .,.,dis}'$) denotes the efficiency factor of the charge (resp., the discharge) initial intercalation fractions. If no acceptable solution is found with any of the floating parameters, the data point is rejected from the dataset. The new dataset contains 93,000 pairs of $(\theta, x)$. After retraining the CNPE, the LLI and $\rm{LAM}_{\rm PE}$ prediction are computed and shown in Fig.~\ref{fig:LLILAM_licons}.

Compared to Fig.~\ref{fig:LLILAM},  $\rm{LLI}_{\rm ch}$ and $\rm{LLI}_{\rm dis}$ are significantly closer in Fig.~\ref{fig:LLILAM_licons} after enforcing Li conservation within a particular cycle. As the cell ages, both LLI metrics start departing from each other but remain within 1-2\% of LLI. 

\subsection{Tractability in high-dimensional parameter spaces}
\label{sec:tractability}

As compared to Bayesian calibration, NPE is less amenable to physics-informed training and can be expected to be more data-hungry. The objective of the present section is to quantify the data requirements of NPE and understand whether the method is still applicable as the number of parameter $N_{\theta}$ increases. We consider the parameter inference problem using voltage from $\rm{C}/2$ discharge data. The parameters inferred are inputs of a P2D model. We consider $N_{\theta} = \{ 6, 16, 27\}$. The parameters of the 6 dimensional cases are the ones shown in Tab.~\ref{tab:priorcompds}. For the higher-dimensional case, the parameter priors are provided in \ref{app:priorhighdim}. 

A performance metric is defined as the sum of the parameter and coverage error ($\varepsilon_{\rm PE} + \varepsilon_{\rm CE}$) to combine the effect of mean parameter prediction and coverage. For each $N_{\theta}$-dimensional case, the size of the training data is varied until the performance metric converges. The same test set is used and always contains 10,000 samples. In each case, the training is done for 350,000 steps by adjusting the total number of epochs. 

\begin{figure}
    \centering \includegraphics[width=0.49\textwidth]{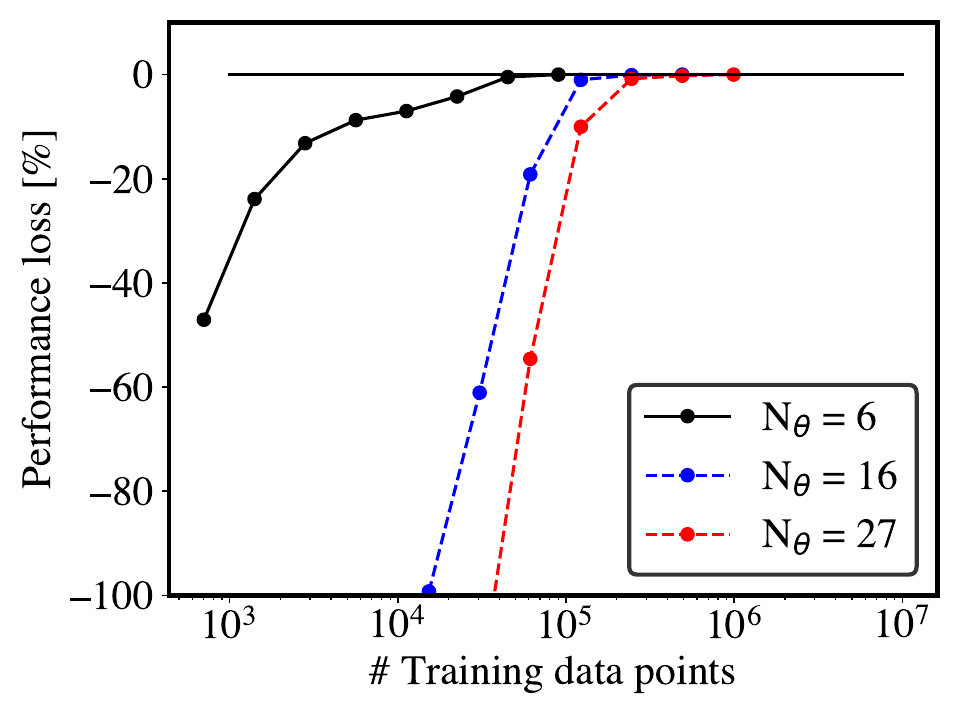}
    \includegraphics[width=0.49\textwidth]{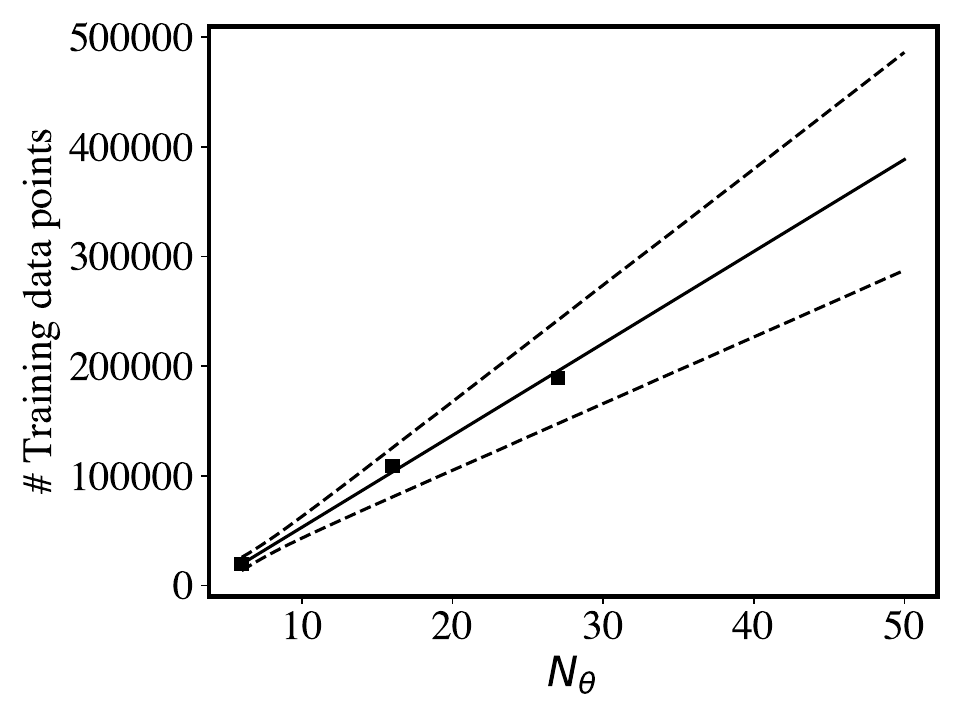}
    \caption{Left: performance metric loss with respect to the training data size for the 6-dimensional, 16-dimensional and 27-dimensional cases. Right: number of training data point required to lose at most 5\% for the performance metric. Results at 6,16 and 27 dimensions (\mythicksquare{black}{black}) are super-imposed with a linear trend (\mythickline{black}) and associated multiplicative uncertainty (\mythickdashedline{black}).}
    \label{fig:highdim}
\end{figure}

For the 6-dimensional ($N_{\theta}$ = 6) case, a partial hyperparameter search is conducted for every new training dataset by adjusting the number of channels and the width of the fully connected layers. In general, as the number of training data points decreases, the optimal number of trainable parameters also decreases. For the 16 and 27 dimensional cases, the CNPE architecture is held fixed to the one described in \ref{app:hypercnpe}. Figure~\ref{fig:highdim} (left) shows how the performance metric is impacted as the number of training data points decreases. In all cases, it can be seen that more data means higher performance. The drop-off of performance with respect to training data is steeper for the 16 and 27 dimensional cases as compared to the 6-dimensional case, which is likely the result of performing hyperparameter optimization only in the 6-dimensional case.

Using a spline interpolation, we compute the number of training data points needed to achieve at most a 5\% drop in performance compared to the best performing case. This number is an indication of how much data is needed to reliably train the CNPE. To extract a trend across parameter dimensions, we assume a linear increase in data requirements with respect to $N_{\theta}$. We also calibrate a multiplicative error to reflect that fixing hyperparameters likely has a higher effect as $N_{\theta}$ grows. The results are shown in Fig.~\ref{fig:highdim} (right). For $N_{\theta}=6$, it is clear that 100,000 data points was sufficient. As $N_{\theta}$ increases, the data requirements also grow but are unlikely to reach more than 500,000 data points. Here, we were able to train the 27 parameter case with as much as 1,000,000 data points, which shows that the data requirements are large, but not out of reach. Data requirements could be reduced even further using sequential posterior estimation \cite{papamakarios2016fast} and active sampling methods for simulation-based inference \cite{griesemer2024active}.

\section{Conclusions}
\label{sec:conclusions}

In this paper, we show that Neural Posterior Estimation is a viable pathway to build amortized probabilistic parameter inference models for Li-ion batteries. A key benefit of NPE is that inference computational costs can be driven down to milliseconds, enabling on-board diagnostics or diagnostics of fleet of devices. In general, NPE predicts parameters more reliably than Bayesian calibration, as Bayesian calibration tends to overfit the model parameters to adjust for measurement noise. Conversely, the NPE method provides conservative uncertainty estimates that can lead to significant voltage fit mismatch, resulting in robustness to measurement noise but also sometimes oversimplifying the parameter probability distribution function due to a priori specification of the posterior shape. Improved fit accuracy could come from more expressive posteriors~\cite{goldwyn2025multidimensional} or flow-based methods~\cite{papamakarios2021normalizing,lipman2022flow,zhang2026discovery}, which is the objective of future work.  When deployed on experimental datasets, reasonably accurate predictions can be obtained for typical degradation mechanisms. However, in general, NPE could also be over-confidently wrong, especially towards the end of the battery aging process. This suggests that some parameters are erroneously assumed to be fixed, or that the underlying physics-based model is inadequate. The first case could be addressed at the condition of using sufficient data. The second requires a quantification of the physics-based model error. 

From a practical viewpoint, the NPE method described here offers a reliable method to efficiently diagnose internal Li-ion battery model parameters and determine the confidence of these parameter estimates while being computationally efficient on inference.  Importantly, this method requires 1) a predetermined protocol, 2) a physics-based model, and 3) computational resources to generate sufficient realizations of the model to provide sufficient training data to explore the parameter space.  If these criteria are met, the trained NPE method can be used in a cycle-by-cycle fashion to determine how internal battery parameters evolve as the cell ages.  Once trained, explainability tools can be used (e.g., SHAP) to determine the parts of the voltage trace that were most indicative of a particular parameter (e.g., the end-of-charge shape is particularly sensitive to the cathode active material volume fraction).  Furthermore, the NPE method can extend to a significant number of model parameters (i.e., $N_{\theta} = 27$) given enough training data is provided (e.g., $\approx$200,000 for 27 inferred model parameters).

\section*{Acknowledgments}

This work is authored in part by the National Laboratory of the Rockies, operated under Contract No. DE-AC36-08GO28308, and in part by Idaho National Laboratory, operated under Contract No. DE-AC07-05ID14517. Funding is provided by the U.S. DOE, Transportation Technology Office under the Machine Learning for Accelerated Life Prediction and Cell Design (pbML) Program, program manager Simon Thompson. The research was performed using computational resources sponsored by the Department of Energy's Office of Critical Minerals and Energy Innovation and located at the National Laboratory of the Rockies. The views expressed in the article do not necessarily represent the views of the DOE or the U.S. Government. The U.S. Government retains and the publisher, by accepting the article for publication, acknowledges that the U.S. Government retains a nonexclusive, paid-up, irrevocable, worldwide license to publish or reproduce the published form of this work, or allow others to do so, for U.S. Government purposes.

%\bibliography{references}

\begin{thebibliography}{56}
\expandafter\ifx\csname natexlab\endcsname\relax\def\natexlab#1{#1}\fi
\providecommand{\url}[1]{\texttt{#1}}
\providecommand{\href}[2]{#2}
\providecommand{\path}[1]{#1}
\providecommand{\DOIprefix}{doi:}
\providecommand{\ArXivprefix}{arXiv:}
\providecommand{\URLprefix}{URL: }
\providecommand{\Pubmedprefix}{pmid:}
\providecommand{\doi}[1]{\href{http://dx.doi.org/#1}{\path{#1}}}
\providecommand{\Pubmed}[1]{\href{pmid:#1}{\path{#1}}}
\providecommand{\bibinfo}[2]{#2}
\ifx\xfnm\relax \def\xfnm[#1]{\unskip,\space#1}\fi
%Type = Article
\bibitem[{Weddle et~al.(2023)Weddle, Kim, Chen, Yi, Gasper, Colclasure, Smith, Gering, Tanim, and Dufek}]{weddle2023battery}
\bibinfo{author}{P.~J. Weddle}, \bibinfo{author}{S.~Kim}, \bibinfo{author}{B.-R. Chen}, \bibinfo{author}{Z.~Yi}, \bibinfo{author}{P.~Gasper}, \bibinfo{author}{A.~M. Colclasure}, \bibinfo{author}{K.~Smith}, \bibinfo{author}{K.~L. Gering}, \bibinfo{author}{T.~R. Tanim}, \bibinfo{author}{E.~J. Dufek},
\newblock \bibinfo{title}{Battery state-of-health diagnostics during fast cycling using physics-informed deep-learning},
\newblock \bibinfo{journal}{Journal of Power Sources} \bibinfo{volume}{585} (\bibinfo{year}{2023}) \bibinfo{pages}{233582}.
%Type = Article
\bibitem[{Kim et~al.(2022)Kim, Yi, Chen, Tanim, and Dufek}]{kim2022rapid}
\bibinfo{author}{S.~Kim}, \bibinfo{author}{Z.~Yi}, \bibinfo{author}{B.-R. Chen}, \bibinfo{author}{T.~R. Tanim}, \bibinfo{author}{E.~J. Dufek},
\newblock \bibinfo{title}{Rapid failure mode classification and quantification in batteries: A deep learning modeling framework},
\newblock \bibinfo{journal}{Energy Storage Materials} \bibinfo{volume}{45} (\bibinfo{year}{2022}) \bibinfo{pages}{1002--1011}.
%Type = Article
\bibitem[{Duan et~al.(2022)Duan, Liu, Agar, and Jin}]{duan2022parameter}
\bibinfo{author}{X.~Duan}, \bibinfo{author}{F.~Liu}, \bibinfo{author}{E.~Agar}, \bibinfo{author}{X.~Jin},
\newblock \bibinfo{title}{Parameter identification of lithium-ion batteries by coupling electrochemical impedance spectroscopy with a physics-based model},
\newblock \bibinfo{journal}{Journal of The Electrochemical Society} \bibinfo{volume}{169} (\bibinfo{year}{2022}) \bibinfo{pages}{040561}.
%Type = Article
\bibitem[{Li et~al.(2022)Li, Demir, Cao, J{\"o}st, Ringbeck, Junker, and Sauer}]{li2022data}
\bibinfo{author}{W.~Li}, \bibinfo{author}{I.~Demir}, \bibinfo{author}{D.~Cao}, \bibinfo{author}{D.~J{\"o}st}, \bibinfo{author}{F.~Ringbeck}, \bibinfo{author}{M.~Junker}, \bibinfo{author}{D.~U. Sauer},
\newblock \bibinfo{title}{Data-driven systematic parameter identification of an electrochemical model for lithium-ion batteries with artificial intelligence},
\newblock \bibinfo{journal}{Energy Storage Materials} \bibinfo{volume}{44} (\bibinfo{year}{2022}) \bibinfo{pages}{557--570}.
%Type = Article
\bibitem[{Andersson et~al.(2022)Andersson, Streb, Ko, Klass, Klett, Ekstr{\"o}m, Johansson, and Lindbergh}]{andersson2022parametrization}
\bibinfo{author}{M.~Andersson}, \bibinfo{author}{M.~Streb}, \bibinfo{author}{J.~Y. Ko}, \bibinfo{author}{V.~L. Klass}, \bibinfo{author}{M.~Klett}, \bibinfo{author}{H.~Ekstr{\"o}m}, \bibinfo{author}{M.~Johansson}, \bibinfo{author}{G.~Lindbergh},
\newblock \bibinfo{title}{{Parametrization of physics-based battery models from input--output data: A review of methodology and current research}},
\newblock \bibinfo{journal}{Journal of Power Sources} \bibinfo{volume}{521} (\bibinfo{year}{2022}) \bibinfo{pages}{230859}.
%Type = Article
\bibitem[{Dufek et~al.(2022)Dufek, Abraham, Bloom, Chen, Chinnam, Colclasure, Gering, Keyser, Kim, Mai et~al.}]{dufek2022developing}
\bibinfo{author}{E.~J. Dufek}, \bibinfo{author}{D.~P. Abraham}, \bibinfo{author}{I.~Bloom}, \bibinfo{author}{B.-R. Chen}, \bibinfo{author}{P.~R. Chinnam}, \bibinfo{author}{A.~M. Colclasure}, \bibinfo{author}{K.~L. Gering}, \bibinfo{author}{M.~Keyser}, \bibinfo{author}{S.~Kim}, \bibinfo{author}{W.~Mai}, et~al.,
\newblock \bibinfo{title}{{Developing extreme fast charge battery protocols--A review spanning materials to systems}},
\newblock \bibinfo{journal}{Journal of Power Sources} \bibinfo{volume}{526} (\bibinfo{year}{2022}) \bibinfo{pages}{231129}.
%Type = Article
\bibitem[{Guittet et~al.(2024)Guittet, Gasper, Shirk, Mitchell, Gilleran, Bonnema, Smith, Mishra, and Mann}]{guittet2024levelized}
\bibinfo{author}{D.~Guittet}, \bibinfo{author}{P.~Gasper}, \bibinfo{author}{M.~Shirk}, \bibinfo{author}{M.~Mitchell}, \bibinfo{author}{M.~Gilleran}, \bibinfo{author}{E.~Bonnema}, \bibinfo{author}{K.~Smith}, \bibinfo{author}{P.~Mishra}, \bibinfo{author}{M.~Mann},
\newblock \bibinfo{title}{{Levelized cost of charging of extreme fast charging with stationary LMO/LTO batteries}},
\newblock \bibinfo{journal}{Journal of Energy Storage} \bibinfo{volume}{82} (\bibinfo{year}{2024}) \bibinfo{pages}{110568}.
%Type = Article
\bibitem[{Reniers et~al.(2021)Reniers, Mulder, and Howey}]{reniers2021unlocking}
\bibinfo{author}{J.~M. Reniers}, \bibinfo{author}{G.~Mulder}, \bibinfo{author}{D.~A. Howey},
\newblock \bibinfo{title}{Unlocking extra value from grid batteries using advanced models},
\newblock \bibinfo{journal}{Journal of power sources} \bibinfo{volume}{487} (\bibinfo{year}{2021}) \bibinfo{pages}{229355}.
%Type = Article
\bibitem[{Zhang et~al.(2026)Zhang, Zhang, Yi, Ren, Jiao, Bai, Jiang, and Song}]{zhang2026discovery}
\bibinfo{author}{J.~Zhang}, \bibinfo{author}{Y.~Zhang}, \bibinfo{author}{B.~Yi}, \bibinfo{author}{Y.~Ren}, \bibinfo{author}{Q.~Jiao}, \bibinfo{author}{H.~Bai}, \bibinfo{author}{W.~Jiang}, \bibinfo{author}{Z.~Song},
\newblock \bibinfo{title}{Discovery learning predicts battery cycle life from minimal experiments},
\newblock \bibinfo{journal}{Nature} \bibinfo{volume}{650} (\bibinfo{year}{2026}) \bibinfo{pages}{110--115}.
%Type = Article
\bibitem[{Hassanaly et~al.(2024)Hassanaly, Weddle, King, De, Doostan, Randall, Dufek, Colclasure, and Smith}]{hassanaly2024pinn2}
\bibinfo{author}{M.~Hassanaly}, \bibinfo{author}{P.~J. Weddle}, \bibinfo{author}{R.~N. King}, \bibinfo{author}{S.~De}, \bibinfo{author}{A.~Doostan}, \bibinfo{author}{C.~R. Randall}, \bibinfo{author}{E.~J. Dufek}, \bibinfo{author}{A.~M. Colclasure}, \bibinfo{author}{K.~Smith},
\newblock \bibinfo{title}{{PINN surrogate of Li-ion battery models for parameter inference, Part II: Regularization and application of the pseudo-2D model}},
\newblock \bibinfo{journal}{Journal of Energy Storage} \bibinfo{volume}{98} (\bibinfo{year}{2024}) \bibinfo{pages}{113104}.
%Type = Article
\bibitem[{Reddy et~al.(2019)Reddy, Scharrer, Pichler, Watzenig, and Dulikravich}]{reddy2019accelerating}
\bibinfo{author}{S.~R. Reddy}, \bibinfo{author}{M.~K. Scharrer}, \bibinfo{author}{F.~Pichler}, \bibinfo{author}{D.~Watzenig}, \bibinfo{author}{G.~S. Dulikravich},
\newblock \bibinfo{title}{{Accelerating parameter estimation in Doyle--Fuller--Newman model for lithium-ion batteries}},
\newblock \bibinfo{journal}{COMPEL-The international journal for computation and mathematics in electrical and electronic engineering} \bibinfo{volume}{38} (\bibinfo{year}{2019}) \bibinfo{pages}{1533--1544}.
%Type = Article
\bibitem[{Santhanagopalan et~al.(2006)Santhanagopalan, Guo, Ramadass, and White}]{santhanagopalan2006review}
\bibinfo{author}{S.~Santhanagopalan}, \bibinfo{author}{Q.~Guo}, \bibinfo{author}{P.~Ramadass}, \bibinfo{author}{R.~E. White},
\newblock \bibinfo{title}{Review of models for predicting the cycling performance of lithium ion batteries},
\newblock \bibinfo{journal}{Journal of power sources} \bibinfo{volume}{156} (\bibinfo{year}{2006}) \bibinfo{pages}{620--628}.
%Type = Article
\bibitem[{Doyle et~al.(1993)Doyle, Fuller, and Newman}]{doyle1993modeling}
\bibinfo{author}{M.~Doyle}, \bibinfo{author}{T.~F. Fuller}, \bibinfo{author}{J.~Newman},
\newblock \bibinfo{title}{Modeling of galvanostatic charge and discharge of the lithium/polymer/insertion cell},
\newblock \bibinfo{journal}{J. Electrochem. Soc} \bibinfo{volume}{140} (\bibinfo{year}{1993}) \bibinfo{pages}{1526}.
%Type = Article
\bibitem[{Fuller et~al.(1994)Fuller, Doyle, and Newman}]{fuller1994relaxation}
\bibinfo{author}{T.~F. Fuller}, \bibinfo{author}{M.~Doyle}, \bibinfo{author}{J.~Newman},
\newblock \bibinfo{title}{Relaxation phenomena in lithium-ion-insertion cells},
\newblock \bibinfo{journal}{Journal of the Electrochemical Society} \bibinfo{volume}{141} (\bibinfo{year}{1994}) \bibinfo{pages}{982}.
%Type = Article
\bibitem[{Ramadesigan et~al.(2011)Ramadesigan, Chen, Burns, Boovaragavan, Braatz, and Subramanian}]{ramadesigan2011parameter}
\bibinfo{author}{V.~Ramadesigan}, \bibinfo{author}{K.~Chen}, \bibinfo{author}{N.~A. Burns}, \bibinfo{author}{V.~Boovaragavan}, \bibinfo{author}{R.~D. Braatz}, \bibinfo{author}{V.~R. Subramanian},
\newblock \bibinfo{title}{Parameter estimation and capacity fade analysis of lithium-ion batteries using reformulated models},
\newblock \bibinfo{journal}{Journal of the Electrochemical society} \bibinfo{volume}{158} (\bibinfo{year}{2011}) \bibinfo{pages}{A1048}.
%Type = Article
\bibitem[{Yu et~al.(2025)Yu, Zhang, Zhang, and Yang}]{yu2025state}
\bibinfo{author}{H.~Yu}, \bibinfo{author}{H.~Zhang}, \bibinfo{author}{Z.~Zhang}, \bibinfo{author}{S.~Yang},
\newblock \bibinfo{title}{{State estimation of lithium-ion batteries via physics-machine learning combined methods: A methodological review and future perspectives}},
\newblock \bibinfo{journal}{ETransportation}  (\bibinfo{year}{2025}) \bibinfo{pages}{100420}.
%Type = Article
\bibitem[{Hassanaly et~al.(2024)Hassanaly, Weddle, King, De, Doostan, Randall, Dufek, Colclasure, and Smith}]{hassanaly2024pinn1}
\bibinfo{author}{M.~Hassanaly}, \bibinfo{author}{P.~J. Weddle}, \bibinfo{author}{R.~N. King}, \bibinfo{author}{S.~De}, \bibinfo{author}{A.~Doostan}, \bibinfo{author}{C.~R. Randall}, \bibinfo{author}{E.~J. Dufek}, \bibinfo{author}{A.~M. Colclasure}, \bibinfo{author}{K.~Smith},
\newblock \bibinfo{title}{{PINN surrogate of Li-ion battery models for parameter inference, Part I: Implementation and multi-fidelity hierarchies for the single-particle model}},
\newblock \bibinfo{journal}{Journal of Energy Storage} \bibinfo{volume}{98} (\bibinfo{year}{2024}) \bibinfo{pages}{113103}.
%Type = Article
\bibitem[{Li et~al.(2024)Li, Li, Yuan, and Zhang}]{li2024deep}
\bibinfo{author}{J.~Li}, \bibinfo{author}{X.~Li}, \bibinfo{author}{X.~Yuan}, \bibinfo{author}{Y.~Zhang},
\newblock \bibinfo{title}{Deep learning method for online parameter identification of lithium-ion batteries using electrochemical synthetic data},
\newblock \bibinfo{journal}{Energy Storage Materials} \bibinfo{volume}{72} (\bibinfo{year}{2024}) \bibinfo{pages}{103697}.
%Type = Article
\bibitem[{Ko et~al.(2024)Ko, Lu, Chen, and Chen}]{ko2024using}
\bibinfo{author}{C.-J. Ko}, \bibinfo{author}{C.-W. Lu}, \bibinfo{author}{K.-C. Chen}, \bibinfo{author}{C.-H. Chen},
\newblock \bibinfo{title}{Using partial discharge data to identify highly sensitive electrochemical parameters of aged lithium-ion batteries},
\newblock \bibinfo{journal}{Energy Storage Materials} \bibinfo{volume}{71} (\bibinfo{year}{2024}) \bibinfo{pages}{103665}.
%Type = Article
\bibitem[{Lenzi et~al.(2023)Lenzi, Bessac, Rudi, and Stein}]{lenzi2023neural}
\bibinfo{author}{A.~Lenzi}, \bibinfo{author}{J.~Bessac}, \bibinfo{author}{J.~Rudi}, \bibinfo{author}{M.~L. Stein},
\newblock \bibinfo{title}{Neural networks for parameter estimation in intractable models},
\newblock \bibinfo{journal}{Computational Statistics \& Data Analysis} \bibinfo{volume}{185} (\bibinfo{year}{2023}) \bibinfo{pages}{107762}.
%Type = Article
\bibitem[{Brendel et~al.(2025)Brendel, Mele, Rosskopf, Katra{\v{s}}nik, and Lorentz}]{brendel2025parametrized}
\bibinfo{author}{P.~Brendel}, \bibinfo{author}{I.~Mele}, \bibinfo{author}{A.~Rosskopf}, \bibinfo{author}{T.~Katra{\v{s}}nik}, \bibinfo{author}{V.~Lorentz},
\newblock \bibinfo{title}{{Parametrized physics-informed deep operator networks for Design of Experiments applied to Lithium-Ion-Battery cells}},
\newblock \bibinfo{journal}{Journal of Energy Storage} \bibinfo{volume}{128} (\bibinfo{year}{2025}) \bibinfo{pages}{117055}.
%Type = Article
\bibitem[{Rom{\'a}n-Ram{\'\i}rez and Marco(2022)}]{roman2022design}
\bibinfo{author}{L.~A. Rom{\'a}n-Ram{\'\i}rez}, \bibinfo{author}{J.~Marco},
\newblock \bibinfo{title}{{Design of experiments applied to lithium-ion batteries: A literature review}},
\newblock \bibinfo{journal}{Applied Energy} \bibinfo{volume}{320} (\bibinfo{year}{2022}) \bibinfo{pages}{119305}.
%Type = Article
\bibitem[{Wang et~al.(2023)Wang, Zhou, Zhang, Sun, Shi, and Huang}]{wang2023parameter}
\bibinfo{author}{Z.~Wang}, \bibinfo{author}{X.~Zhou}, \bibinfo{author}{W.~Zhang}, \bibinfo{author}{B.~Sun}, \bibinfo{author}{J.~Shi}, \bibinfo{author}{Q.~Huang},
\newblock \bibinfo{title}{Parameter sensitivity analysis and parameter identifiability analysis of electrochemical model under wide discharge rate},
\newblock \bibinfo{journal}{Journal of Energy Storage} \bibinfo{volume}{68} (\bibinfo{year}{2023}) \bibinfo{pages}{107788}.
%Type = Article
\bibitem[{Nascimento et~al.(2023)Nascimento, Viana, Corbetta, and Kulkarni}]{nascimento2023framework}
\bibinfo{author}{R.~G. Nascimento}, \bibinfo{author}{F.~A. Viana}, \bibinfo{author}{M.~Corbetta}, \bibinfo{author}{C.~S. Kulkarni},
\newblock \bibinfo{title}{{A framework for Li-ion battery prognosis based on hybrid Bayesian physics-informed neural networks}},
\newblock \bibinfo{journal}{Scientific Reports} \bibinfo{volume}{13} (\bibinfo{year}{2023}) \bibinfo{pages}{13856}.
%Type = Article
\bibitem[{Kim et~al.(2023)Kim, Kim, Choi, and Choi}]{kim2023bayesian}
\bibinfo{author}{S.~Kim}, \bibinfo{author}{S.~Kim}, \bibinfo{author}{Y.~Y. Choi}, \bibinfo{author}{J.-I. Choi},
\newblock \bibinfo{title}{Bayesian parameter identification in electrochemical model for lithium-ion batteries},
\newblock \bibinfo{journal}{Journal of Energy Storage} \bibinfo{volume}{71} (\bibinfo{year}{2023}) \bibinfo{pages}{108129}.
%Type = Article
\bibitem[{Aitio et~al.(2020)Aitio, Marquis, Ascencio, and Howey}]{aitio2020bayesian}
\bibinfo{author}{A.~Aitio}, \bibinfo{author}{S.~G. Marquis}, \bibinfo{author}{P.~Ascencio}, \bibinfo{author}{D.~Howey},
\newblock \bibinfo{title}{{Bayesian parameter estimation applied to the Li-ion battery single particle model with electrolyte dynamics}},
\newblock \bibinfo{journal}{IFAC-PapersOnLine} \bibinfo{volume}{53} (\bibinfo{year}{2020}) \bibinfo{pages}{12497--12504}.
%Type = Article
\bibitem[{Bills et~al.(2023)Bills, Fredericks, Sulzer, and Viswanathan}]{bills2023massively}
\bibinfo{author}{A.~Bills}, \bibinfo{author}{L.~Fredericks}, \bibinfo{author}{V.~Sulzer}, \bibinfo{author}{V.~Viswanathan},
\newblock \bibinfo{title}{Massively distributed bayesian analysis of electric aircraft battery degradation},
\newblock \bibinfo{journal}{ACS Energy Letters} \bibinfo{volume}{8} (\bibinfo{year}{2023}) \bibinfo{pages}{3578--3585}.
%Type = Article
\bibitem[{Cranmer et~al.(2020)Cranmer, Brehmer, and Louppe}]{cranmer2020frontier}
\bibinfo{author}{K.~Cranmer}, \bibinfo{author}{J.~Brehmer}, \bibinfo{author}{G.~Louppe},
\newblock \bibinfo{title}{The frontier of simulation-based inference},
\newblock \bibinfo{journal}{Proceedings of the National Academy of Sciences} \bibinfo{volume}{117} (\bibinfo{year}{2020}) \bibinfo{pages}{30055--30062}.
%Type = Article
\bibitem[{Deistler et~al.(2025)Deistler, Boelts, Steinbach, Moss, Moreau, Gloeckler, Rodrigues, Linhart, Lappalainen, Miller et~al.}]{deistler2025simulation}
\bibinfo{author}{M.~Deistler}, \bibinfo{author}{J.~Boelts}, \bibinfo{author}{P.~Steinbach}, \bibinfo{author}{G.~Moss}, \bibinfo{author}{T.~Moreau}, \bibinfo{author}{M.~Gloeckler}, \bibinfo{author}{P.~L. Rodrigues}, \bibinfo{author}{J.~Linhart}, \bibinfo{author}{J.~K. Lappalainen}, \bibinfo{author}{B.~K. Miller}, et~al.,
\newblock \bibinfo{title}{{Simulation-Based Inference: A Practical Guide}},
\newblock \bibinfo{journal}{arXiv preprint arXiv:2508.12939}  (\bibinfo{year}{2025}).
%Type = Article
\bibitem[{Hoffman et~al.(2014)Hoffman, Gelman et~al.}]{hoffman2014no}
\bibinfo{author}{M.~D. Hoffman}, \bibinfo{author}{A.~Gelman}, et~al.,
\newblock \bibinfo{title}{{The No-U-Turn sampler: adaptively setting path lengths in Hamiltonian Monte Carlo}},
\newblock \bibinfo{journal}{J. Mach. Learn. Res.} \bibinfo{volume}{15} (\bibinfo{year}{2014}) \bibinfo{pages}{1593--1623}.
%Type = Article
\bibitem[{Nemeth and Fearnhead(2021)}]{nemeth2021stochastic}
\bibinfo{author}{C.~Nemeth}, \bibinfo{author}{P.~Fearnhead},
\newblock \bibinfo{title}{{Stochastic gradient Markov Chain Monte Carlo}},
\newblock \bibinfo{journal}{Journal of the American Statistical Association} \bibinfo{volume}{116} (\bibinfo{year}{2021}) \bibinfo{pages}{433--450}.
%Type = Article
\bibitem[{Kullback and Leibler(1951)}]{kullback1951information}
\bibinfo{author}{S.~Kullback}, \bibinfo{author}{R.~A. Leibler},
\newblock \bibinfo{title}{On information and sufficiency},
\newblock \bibinfo{journal}{The annals of mathematical statistics} \bibinfo{volume}{22} (\bibinfo{year}{1951}) \bibinfo{pages}{79--86}.
%Type = Article
\bibitem[{Blei et~al.(2017)Blei, Kucukelbir, and McAuliffe}]{blei2017variational}
\bibinfo{author}{D.~M. Blei}, \bibinfo{author}{A.~Kucukelbir}, \bibinfo{author}{J.~D. McAuliffe},
\newblock \bibinfo{title}{{Variational inference: A review for statisticians}},
\newblock \bibinfo{journal}{Journal of the American statistical Association} \bibinfo{volume}{112} (\bibinfo{year}{2017}) \bibinfo{pages}{859--877}.
%Type = Article
\bibitem[{Papamakarios and Murray(2016)}]{papamakarios2016fast}
\bibinfo{author}{G.~Papamakarios}, \bibinfo{author}{I.~Murray},
\newblock \bibinfo{title}{Fast $\varepsilon$-free inference of simulation models with bayesian conditional density estimation},
\newblock \bibinfo{journal}{Advances in neural information processing systems} \bibinfo{volume}{29} (\bibinfo{year}{2016}).
%Type = Article
\bibitem[{Braman et~al.(2013)Braman, Oliver, and Raman}]{braman2013bayesian}
\bibinfo{author}{K.~Braman}, \bibinfo{author}{T.~Oliver}, \bibinfo{author}{V.~Raman},
\newblock \bibinfo{title}{Bayesian analysis of syngas chemistry models},
\newblock \bibinfo{journal}{Combust. Theory Model} \bibinfo{volume}{17} (\bibinfo{year}{2013}) \bibinfo{pages}{858--887}.
%Type = Article
\bibitem[{Hassanaly et~al.(2025)Hassanaly, Parra-Alvarez, Rahimi, Municchi, and Sitaraman}]{hassanaly2025bayesian}
\bibinfo{author}{M.~Hassanaly}, \bibinfo{author}{J.~M. Parra-Alvarez}, \bibinfo{author}{M.~J. Rahimi}, \bibinfo{author}{F.~Municchi}, \bibinfo{author}{H.~Sitaraman},
\newblock \bibinfo{title}{{Bayesian calibration of bubble size dynamics applied to CO$_2$ gas fermenters}},
\newblock \bibinfo{journal}{Chemical Engineering Research and Design} \bibinfo{volume}{215} (\bibinfo{year}{2025}) \bibinfo{pages}{312--328}.
%Type = Article
\bibitem[{Wildberger et~al.(2023)Wildberger, Dax, Buchholz, Green, Macke, and Sch{\"o}lkopf}]{wildberger2023flow}
\bibinfo{author}{J.~Wildberger}, \bibinfo{author}{M.~Dax}, \bibinfo{author}{S.~Buchholz}, \bibinfo{author}{S.~Green}, \bibinfo{author}{J.~H. Macke}, \bibinfo{author}{B.~Sch{\"o}lkopf},
\newblock \bibinfo{title}{Flow matching for scalable simulation-based inference},
\newblock \bibinfo{journal}{Advances in Neural Information Processing Systems} \bibinfo{volume}{36} (\bibinfo{year}{2023}) \bibinfo{pages}{16837--16864}.
%Type = Article
\bibitem[{Hassanaly et~al.(2022)Hassanaly, Glaws, Stengel, and King}]{hassanaly2022adversarial}
\bibinfo{author}{M.~Hassanaly}, \bibinfo{author}{A.~Glaws}, \bibinfo{author}{K.~Stengel}, \bibinfo{author}{R.~N. King},
\newblock \bibinfo{title}{Adversarial sampling of unknown and high-dimensional conditional distributions},
\newblock \bibinfo{journal}{Journal of Computational Physics} \bibinfo{volume}{450} (\bibinfo{year}{2022}) \bibinfo{pages}{110853}.
%Type = Misc
\bibitem[{Randall(2025)}]{randall2025bmlite}
\bibinfo{author}{C.~R. Randall}, \bibinfo{title}{{BATMODS-lite: Packaged battery models and material properties [SWR-25-108]}}, \bibinfo{year}{2025}. \URLprefix \url{github.com/NatLabRockies/batmods-lite}. \DOIprefix\doi{10.11578/dc.20260114.1}.
%Type = Article
\bibitem[{Chen et~al.(2022)Chen, Walker, Kim, Kunz, Tanim, and Dufek}]{chen2022battery}
\bibinfo{author}{B.-R. Chen}, \bibinfo{author}{C.~M. Walker}, \bibinfo{author}{S.~Kim}, \bibinfo{author}{M.~R. Kunz}, \bibinfo{author}{T.~R. Tanim}, \bibinfo{author}{E.~J. Dufek},
\newblock \bibinfo{title}{{Battery aging mode identification across NMC compositions and designs using machine learning}},
\newblock \bibinfo{journal}{Joule} \bibinfo{volume}{6} (\bibinfo{year}{2022}) \bibinfo{pages}{2776--2793}.
%Type = Article
\bibitem[{Villalobos et~al.(2025)Villalobos, Rudi, and Mang}]{villalobos2025neural}
\bibinfo{author}{G.~Villalobos}, \bibinfo{author}{J.~Rudi}, \bibinfo{author}{A.~Mang},
\newblock \bibinfo{title}{{Neural Networks for Bayesian Inverse Problems Governed by a Nonlinear ODE}},
\newblock \bibinfo{journal}{arXiv preprint arXiv:2510.14197}  (\bibinfo{year}{2025}).
%Type = Article
\bibitem[{Papamakarios et~al.(2021)Papamakarios, Nalisnick, Rezende, Mohamed, and Lakshminarayanan}]{papamakarios2021normalizing}
\bibinfo{author}{G.~Papamakarios}, \bibinfo{author}{E.~Nalisnick}, \bibinfo{author}{D.~J. Rezende}, \bibinfo{author}{S.~Mohamed}, \bibinfo{author}{B.~Lakshminarayanan},
\newblock \bibinfo{title}{Normalizing flows for probabilistic modeling and inference},
\newblock \bibinfo{journal}{Journal of Machine Learning Research} \bibinfo{volume}{22} (\bibinfo{year}{2021}) \bibinfo{pages}{1--64}.
%Type = Article
\bibitem[{Phan et~al.(2019)Phan, Pradhan, and Jankowiak}]{numpyro}
\bibinfo{author}{D.~Phan}, \bibinfo{author}{N.~Pradhan}, \bibinfo{author}{M.~Jankowiak},
\newblock \bibinfo{title}{{Composable effects for flexible and accelerated probabilistic programming in NumPyro}},
\newblock \bibinfo{journal}{arXiv preprint arXiv:1912.11554}  (\bibinfo{year}{2019}).
%Type = Article
\bibitem[{Angelopoulos and Bates(2021)}]{angelopoulos2021gentle}
\bibinfo{author}{A.~N. Angelopoulos}, \bibinfo{author}{S.~Bates},
\newblock \bibinfo{title}{A gentle introduction to conformal prediction and distribution-free uncertainty quantification},
\newblock \bibinfo{journal}{arXiv preprint arXiv:2107.07511}  (\bibinfo{year}{2021}).
%Type = Article
\bibitem[{Flamary et~al.(2021)Flamary, Courty, Gramfort, Alaya, Boisbunon, Chambon, Chapel, Corenflos, Fatras, Fournier, Gautheron, Gayraud, Janati, Rakotomamonjy, Redko, Rolet, Schutz, Seguy, Sutherland, Tavenard, Tong, and Vayer}]{flamary2021pot}
\bibinfo{author}{R.~Flamary}, \bibinfo{author}{N.~Courty}, \bibinfo{author}{A.~Gramfort}, \bibinfo{author}{M.~Z. Alaya}, \bibinfo{author}{A.~Boisbunon}, \bibinfo{author}{S.~Chambon}, \bibinfo{author}{L.~Chapel}, \bibinfo{author}{A.~Corenflos}, \bibinfo{author}{K.~Fatras}, \bibinfo{author}{N.~Fournier}, \bibinfo{author}{L.~Gautheron}, \bibinfo{author}{N.~T. Gayraud}, \bibinfo{author}{H.~Janati}, \bibinfo{author}{A.~Rakotomamonjy}, \bibinfo{author}{I.~Redko}, \bibinfo{author}{A.~Rolet}, \bibinfo{author}{A.~Schutz}, \bibinfo{author}{V.~Seguy}, \bibinfo{author}{D.~J. Sutherland}, \bibinfo{author}{R.~Tavenard}, \bibinfo{author}{A.~Tong}, \bibinfo{author}{T.~Vayer},
\newblock \bibinfo{title}{{POT: Python Optimal Transport}},
\newblock \bibinfo{journal}{Journal of Machine Learning Research} \bibinfo{volume}{22} (\bibinfo{year}{2021}) \bibinfo{pages}{1--8}. \URLprefix \url{http://jmlr.org/papers/v22/20-451.html}.
%Type = Inproceedings
\bibitem[{Bonneel et~al.(2011)Bonneel, Van De~Panne, Paris, and Heidrich}]{bonneel2011displacement}
\bibinfo{author}{N.~Bonneel}, \bibinfo{author}{M.~Van De~Panne}, \bibinfo{author}{S.~Paris}, \bibinfo{author}{W.~Heidrich},
\newblock \bibinfo{title}{{Displacement interpolation using Lagrangian mass transport}},
\newblock in: \bibinfo{booktitle}{Proceedings of the 2011 SIGGRAPH Asia conference}, \bibinfo{year}{2011}, pp. \bibinfo{pages}{1--12}.
%Type = Article
\bibitem[{Perr-Sauer et~al.(2025)Perr-Sauer, Ugirumurera, Gafur, Bensen, Nguyen, Paul, Severino, Nag, Vijayshankar, Gasper et~al.}]{perr2025applications}
\bibinfo{author}{J.~Perr-Sauer}, \bibinfo{author}{J.~Ugirumurera}, \bibinfo{author}{J.~Gafur}, \bibinfo{author}{E.~A. Bensen}, \bibinfo{author}{T.~Nguyen}, \bibinfo{author}{S.~Paul}, \bibinfo{author}{J.~Severino}, \bibinfo{author}{A.~Nag}, \bibinfo{author}{S.~Vijayshankar}, \bibinfo{author}{P.~Gasper}, et~al.,
\newblock \bibinfo{title}{Applications of explainable artificial intelligence in renewable energy research},
\newblock \bibinfo{journal}{Energy Reports} \bibinfo{volume}{14} (\bibinfo{year}{2025}) \bibinfo{pages}{2217--2235}.
%Type = Article
\bibitem[{Lundberg and Lee(2017)}]{lundberg2017unified}
\bibinfo{author}{S.~M. Lundberg}, \bibinfo{author}{S.-I. Lee},
\newblock \bibinfo{title}{A unified approach to interpreting model predictions},
\newblock \bibinfo{journal}{Advances in neural information processing systems} \bibinfo{volume}{30} (\bibinfo{year}{2017}).
%Type = Inproceedings
\bibitem[{Shrikumar et~al.(2017)Shrikumar, Greenside, and Kundaje}]{shrikumar2017learning}
\bibinfo{author}{A.~Shrikumar}, \bibinfo{author}{P.~Greenside}, \bibinfo{author}{A.~Kundaje},
\newblock \bibinfo{title}{Learning important features through propagating activation differences},
\newblock in: \bibinfo{booktitle}{International conference on machine learning}, \bibinfo{organization}{PMlR}, \bibinfo{year}{2017}, pp. \bibinfo{pages}{3145--3153}.
%Type = Article
\bibitem[{Griesemer et~al.(2024)Griesemer, Cao, Cui, Osorio, and Liu}]{griesemer2024active}
\bibinfo{author}{S.~Griesemer}, \bibinfo{author}{D.~Cao}, \bibinfo{author}{Z.~Cui}, \bibinfo{author}{C.~Osorio}, \bibinfo{author}{Y.~Liu},
\newblock \bibinfo{title}{Active sequential posterior estimation for sample-efficient simulation-based inference},
\newblock \bibinfo{journal}{Advances in Neural Information Processing Systems} \bibinfo{volume}{37} (\bibinfo{year}{2024}) \bibinfo{pages}{127907--127936}.
%Type = Article
\bibitem[{Goldwyn et~al.(2025)Goldwyn, Krock, Rudi, Getter, and Bessac}]{goldwyn2025multidimensional}
\bibinfo{author}{H.~J. Goldwyn}, \bibinfo{author}{M.~Krock}, \bibinfo{author}{J.~Rudi}, \bibinfo{author}{D.~Getter}, \bibinfo{author}{J.~Bessac},
\newblock \bibinfo{title}{{Multidimensional Distributional Neural Network Output Demonstrated in Super-Resolution of Surface Wind Speed}},
\newblock \bibinfo{journal}{arXiv preprint arXiv:2508.16686}  (\bibinfo{year}{2025}).
%Type = Article
\bibitem[{Lipman et~al.(2022)Lipman, Chen, Ben-Hamu, Nickel, and Le}]{lipman2022flow}
\bibinfo{author}{Y.~Lipman}, \bibinfo{author}{R.~T. Chen}, \bibinfo{author}{H.~Ben-Hamu}, \bibinfo{author}{M.~Nickel}, \bibinfo{author}{M.~Le},
\newblock \bibinfo{title}{Flow matching for generative modeling},
\newblock \bibinfo{journal}{arXiv preprint arXiv:2210.02747}  (\bibinfo{year}{2022}).
%Type = Article
\bibitem[{Sulzer et~al.(2021)Sulzer, Marquis, Timms, Robinson, and Chapman}]{sulzer2021pybamm}
\bibinfo{author}{V.~Sulzer}, \bibinfo{author}{S.~G. Marquis}, \bibinfo{author}{R.~Timms}, \bibinfo{author}{M.~Robinson}, \bibinfo{author}{S.~J. Chapman},
\newblock \bibinfo{title}{{Python Battery Mathematical Modelling (PyBaMM)}},
\newblock \bibinfo{journal}{Journal of Open Research Software} \bibinfo{volume}{9} (\bibinfo{year}{2021}) \bibinfo{pages}{14}. \DOIprefix\doi{10.5334/jors.309}.
%Type = Misc
\bibitem[{Randall(2024)}]{randall2024sksundae}
\bibinfo{author}{C.~R. Randall}, \bibinfo{title}{{scikit-SUNDAE: Python bindings to SUNDIALS differential algebraic equation solvers [SWR-24-137]}}, \bibinfo{year}{2024}. \URLprefix \url{github.com/NatLabRockies/scikit-sundae}. \DOIprefix\doi{10.11578/dc.20241104.3}.
%Type = Article
\bibitem[{Hindmarsh et~al.(2005)Hindmarsh, Brown, Grant, Lee, Serban, Shumaker, and Woodward}]{hindmarsh2005sundials}
\bibinfo{author}{A.~C. Hindmarsh}, \bibinfo{author}{P.~N. Brown}, \bibinfo{author}{K.~E. Grant}, \bibinfo{author}{S.~L. Lee}, \bibinfo{author}{R.~Serban}, \bibinfo{author}{D.~E. Shumaker}, \bibinfo{author}{C.~S. Woodward},
\newblock \bibinfo{title}{{SUNDIALS: Suite of nonlinear and differential/algebraic equation solvers}},
\newblock \bibinfo{journal}{ACM Transactions on Mathematical Software (TOMS)} \bibinfo{volume}{31} (\bibinfo{year}{2005}) \bibinfo{pages}{363--396}.
%Type = Article
\bibitem[{Balos et~al.(2025)Balos, Day, Esclapez, Felden, Gardner, Hassanaly, Reynolds, Rood, Sexton, Wimer et~al.}]{balos2025sundials}
\bibinfo{author}{C.~J. Balos}, \bibinfo{author}{M.~Day}, \bibinfo{author}{L.~Esclapez}, \bibinfo{author}{A.~M. Felden}, \bibinfo{author}{D.~J. Gardner}, \bibinfo{author}{M.~Hassanaly}, \bibinfo{author}{D.~R. Reynolds}, \bibinfo{author}{J.~S. Rood}, \bibinfo{author}{J.~M. Sexton}, \bibinfo{author}{N.~T. Wimer}, et~al.,
\newblock \bibinfo{title}{{SUNDIALS time integrators for exascale applications with many independent systems of ordinary differential equations}},
\newblock \bibinfo{journal}{The International Journal of High Performance Computing Applications} \bibinfo{volume}{39} (\bibinfo{year}{2025}) \bibinfo{pages}{123--146}.

\end{thebibliography}

\appendix

\section{Reference parameter values}
\label{app:bol}

\begin{table}[h!]
\centering
\begin{tabular}{|c | c |} 
 \hline
 Parameter [unit] & Value \\ [0.5ex] 
 \hline
 Nominal capacity [Ah] & $1.89 \times 10^{-2}$ \\
 Cell area [m$^2$] & $1.4\times 10^{-3}$  \\ 
 Initial \ce{Li^+} concentration [kmol~m$^{-3}$] & 1.2 \\ \hline
 Anode thickness [$\mu$ m] & 44 \\
 Anode secondary particle radius [$\mu$ m] & 4 \\
 Anode solid-phase volume fraction [-] & 0.6 \\
 Anode electrolyte volume fraction [-] & 0.4 \\
 Anode carbon-binder-domain volume fraction [-] & 0.0569272237 \\
 Anode charge transfer coefficient [-] & 0.5 \\
 Anode solid-phase Bruggeman factor [-] & 2.0 \\
 Anode liquid-phase Bruggeman factor [-] & 2.0 \\
 Anode maximum solid-phase \ce{Li} concentration [kmol~m$^{-3}$] & 30.53 \\
 Discharge anode initial intercalation fraction [-] & 0.91 \\
 Charge anode initial intercalation fraction [-] & 0.07 \\ \hline
 Separator thickness [$\mu$m] & 20.0 \\
 Separator electrolyte volume fraction [-] & 0.4 \\
 Separator liquid-phase Bruggeman factor [-] & 2.0 \\ \hline
 Cathode thickness [$\mu$m] & 42 \\
 Cathode secondary particle radius [$\mu$m] & 1.8 \\
 Cathode solid-phase volume fraction [-] & 0.6 \\
 Cathode electrolyte volume fraction [-] & 0.4 \\
 Cathode carbon-binder-domain volume fraction [-] & 0.12338 \\
 Cathode charge transfer coefficient [-] & 0.5 \\
 Cathode solid-phase Bruggeman factor [-] & 2.0 \\
 Cathode liquid-phase Bruggeman factor [-] & 2.0 \\
 Cathode maximum solid-phase \ce{Li} concentration [kmol~m$^{-3}$] & 49.6 \\
 Discharge cathode initial intercalation fraction [-] & 0.39 \\
 Charge cathode initial intercalation fraction [-] & 0.89 \\ \hline
\end{tabular}
%\caption{Beginning of life parameters scaled with the coefficient inferred\footnote{Also see \hyperlink{https://github.com/NatLabRockies/batmods-lite/blob/main/src/bmlite/P2D/templates/graphite_nmc532.yaml}{https://github.com/NatLabRockies/batmods-lite/blob/main/src/bmlite/P2D/templates/graphite_nmc532.yaml}}.}
\caption{Reference parameters values. Those are the values scaled with the efficiency factors inferred.} 
\label{tab:bolpar}
\end{table}

The physics-based model parameters for both the single-particle model and the pseudo-2D model are obtained from Weddle et al.~\cite{weddle2023battery} and are shown in Tab.~\ref{tab:bolpar}.  These parameters describe a single-layer pouch cell.  The Li$_x$Ni$_{0.5}$Mn$_{0.3}$Co$_{0.2}$O$_2$ cathode is paired with a graphite anode with Gen 2 (3:7 wt\% EC:EMC) electrolyte.  

\section{Physics-based solver and numerical details}
\label{app:bmlite}

All physics-based simulations were performed using BATMODS-lite~\cite{randall2025bmlite}, an open-source battery modeling framework that provides ready-to-run implementations of the single-particle model (SPM) and pseudo-2D (P2D) model, along with the modular building blocks required for custom model development. The package includes material property relationships for common electrode and electrolyte chemistries (e.g., graphite, NMC532, LFP, and Gen2 electrolyte) and is designed to expand as additional materials are characterized. Although this study focuses on constant-current cycling, BATMODS-lite also supports more complex experimental protocols through a flexible input class that enables constant or dynamic current, voltage, or power control with automatic step transitions triggered by time or state limits. Beyond the included models and properties, the framework includes vectorized finite-volume operators, meshing utilities, and structured indexing routines to help organize the system of equations such that the resulting Jacobian retains a banded or patterned structure, enabling more efficient linear solves. This purely numerical framework provides an alternative to other popular packages that rely on symbolic expressions, e.g., PyBaMM~\cite{sulzer2021pybamm}.

The numerical solvers used in BATMODS-lite are provided by scikit-SUNDAE~\cite{randall2024sksundae}, a Python package that provides bindings to SUNDIALS~\cite{hindmarsh2005sundials, balos2025sundials}. scikit-SUNDAE wraps the CVODE (for ordinary differential equations) and IDA (for differential algebraic equations) solvers, which implement adaptive time-stepping algorithms designed for stiff systems such as electrochemical battery models. Although users define their systems of equations through Python, the underlying solvers execute in compiled C code through the SUNDIALS library, providing high-performance integration with only minimal overhead from Python function calls. The solver interface is intentionally framework-agnostic, allowing users to formulate problems using numerical arrays, symbolic expressions, or compiled functions. Users may also optionally provide an analytic Jacobian (hand-coded or using automatic differentiation libraries), define event functions to monitor or terminate simulations, and select optimized linear solver backends for dense, banded, sparse, or iterative solution methods. In the present work, BATMODS-lite formulates both the SPM and P2D models as systems of differential algebraic equations and solves them using the IDA integrator and a banded linear solver. 

For this study, the SPM and P2D models both use 30 points uniformly distributed throughout the particle radius. The P2D model additionally discretizes the electrode and separator using 32 uniformly spaced points in each domain. With the selected geometry and properties from \ref{app:bol}, a single constant-current charge or discharge protocol can be run in $\approx0.3$~s for the SPM and $\approx1.5$~s for the P2D model on a standard laptop. However, the speed of these models is heavily influenced by the mesh and input parameters. A coarser mesh and changes to the material property expressions can often lead to even faster simulations. Because the training datasets required $10^5$--$10^6$ simulations for each model, data generation was performed in parallel as described in Sec.~\ref{sec:datasets}.

%\crr{Useful details about DAE implementation + C-bindings}

\section{Hyperparameters of the surrogate}
\label{app:hypersurr}

The surrogate results in Sec.~\ref{sec:surr_method} are reported for the best surrogate architecture whose hyper-parameters are shown in Tab.~\ref{tab:Hyperparsurr}. The hidden layers of the surrogate are made of a sequence of fully connected layers whose size are based on the a maximal number of neurons per layer and an odd number of layers. Between each hidden layer, the number of neurons is multiplied by 2 before the middle layer, and divided by 2 after. For example, the optimal architecture has a sequence of hidden layers size $[1024, 2048, 1024]$, which corresponds to a maximal hidden layer size of 2048 and 3 hidden layers. Each training run is done over 200 epochs. An exponential learning rate schedule is used: initially, the learning rate is set to the initial learning rate given by Tab.~\ref{tab:Hyperparsurr}. Over the first 150 epochs, the learning rate decrease exponentially to a value 100 times lower than the initial learning rate. 

The hidden activation layer is chosen to be hyperbolic tangent, consistently with Ref.~\cite{hassanaly2024pinn1}. The final activation layer is a sigmoid that allows for constraining the voltage within predefined limits. In the case of the Gr-NMC532 cells, the voltage must be in the range $[3\rm{V}, 4.1\rm{V}]$. The network is constrained to predict a voltage in the range $[2.5\rm{V}, 4.6\rm{V}]$ so that the beginning and end of discharge points give non-zero gradients to the neural net. 

The hyperparameter values are obtained with a grid search of 36 training runs, and the set of values explored are shown in Tab.~\ref{tab:Hyperparsurr}. %\todo{explain what metric was used and why we need the val data}

\begin{table}[h!]
\centering
\begin{tabular}{|c | c | c|} 
 \hline
 Hyper-parameter & Optimal & Set of values explored \\ [0.5ex] 
 \hline
 Maximal number of neurons per hidden layer & 2048 & $\{512, 1024, 2048\}$ \\ 
 Number of hidden layers & 3 & $\{3, 5, 9\}$ \\
 Initial learning rate & $10^{-3}$ & $\{ 10^{-3}, 10^{-2} \}$ \\
 Batch size & $2^{16}$ & $\{ 2^{16}, 2^{17} \}$ \\[1ex] 
 \hline
\end{tabular}
\caption{Hyperparameters of the surrogate model and associated search space during hyperparameter optimization.}
\label{tab:Hyperparsurr}
\end{table}

\section{Hyperparameters of the CNPE}
\label{app:hypercnpe}
% model 104 from tune_spm_discharge_independent_normal_paper_uniform
%tunes model is /projects/mlbatt/ECS248/spm_discharge_uniform_tuned
Similar to the surrogate, the CNPE results in Sec.~\ref{sec:comparison} are shown for an optimal architecture obtained via hyperparameter tuning. To reduce the number of hyperparameters, all the convolution layers are constrained to use the same number of filters. The kernel size of each filter is fixed to 3. Each convolutional layer uses a LeakyReLU activation \cite{hassanaly2022adversarial} and is followed by a MaxPooling operation. Both fully connected networks that predict the posterior mean and variance, use the same number of hidden layers and number of neurons per layer.  Each of the fully connected network's hidden layers use a hyperbolic tangent activation. The network is trained for 1000 epochs. Similar to the surrogate training, over the first 750 epochs, the learning rate decrease exponentially to a value 100 times lower than the initial learning rate.  

The hyperparameter values are obtained with a grid search of 324 training runs, and the set of values explored are shown in Tab.~\ref{tab:Hyperparcnpe}. Note that compared to Tab.~\ref{tab:Hyperparsurr}, each datapoint is a full discharge curve  (compared to a pointwise voltage value) which explains why the batch size is smaller in the case of CNPE. 

\begin{table}[h!]
\centering
\begin{tabular}{|c | c | c|} 
 \hline
 Hyper-parameter & Optimal & Set of values explored \\ [0.5ex] 
 \hline
 Number of convolutional layer &  2 & $\{2, 4, 6\}$ \\ 
 Number of filters per convolutional layer & 512 & $\{512, 1024, 2048\}$ \\
 Number of fully connected hidden layers & 4 & $\{1, 2, 4\}$ \\
 Number of neurons per hidden layers & 2048 & $\{512, 1024, 2048\}$ \\
 Initial learning rate &  $10^{-4}$ & $\{ 10^{-4}, 10^{-5} \}$ \\
 Batch size & $2^8$ & $\{ 2^8, 2^{9} \}$ \\[1ex] 
 \hline
\end{tabular}
\caption{Hyperparameters of the CNPE model and associated search space during hyperparameter optimization.}
\label{tab:Hyperparcnpe}
\end{table}

\section{Effect of experimental noise on parameter prediction}
\label{app:noise}

The magnitude of the experimental noise is increased by a factor 4 and 8 and the parameter and voltage accuracy metrics are shown in Tab.~\ref{tab:acc4} and Tab.~\ref{tab:acc8}. For consistency, the bounds of the likelihood uncertainty used in the Bayesian calibration are also scaled by a factor 4 and 8 in each case, and a CNPE is retrained for each noise level.  Overall, as the experimental noise increases, the parameter accuracy deteriorates in all cases. This is expected since the noise destroys information about the voltage curves, which become less informative about the physics parameters. Consistently with Sec.~\ref{sec:comparison}, CNPE is equally or even more accurate than Bayesian calibration for the parameters, but leads to higher voltage errors. However, as the noise level increases, the coverage error for Bayesian calibration decreases but always remains higher than for CNPE. This suggests that the poor coverage error for Bayesian calibration in Sec.~\ref{sec:comparison} could be due to Bayesian calibration explaining the noise with the parameters inferred, leading to overconfident but erroneous parameter predictions.  

\begin{table}[h!]
\centering
 \begin{tabular}{ |c|c|c|c|c| } 
        \hline
        Metric &  \multicolumn{2}{|c|}{Bayesian calibration} & \multicolumn{2}{|c|}{CNPE} \\
        \hline
         & Discharge $\rm{C}/2$ & Charge $4\rm{C}$ & Discharge $\rm{C}/2$ & Charge $4\rm{C}$  \\
        \hline
         $\varepsilon_{\rm PE}$   &  $ 8.89 \%$ & $7.29\%$  & \boldsymbol{$ 8.88 \%$} & \boldsymbol{$5.96\%$}\\
         $\varepsilon_{\rm PE, s}$   &  \boldsymbol{$ 11.52 \%$} & $9.36 \%$  & $ 12.18 \%$ & \boldsymbol{$ 8.57\%$} \\ 
          $\varepsilon_{\rm CE}$   &  $ 22.78 \%$  &  $13.45\%$ & \boldsymbol{$ 4.22\%$} &  \boldsymbol{$0.68\%$} \\ 
         $\varepsilon_{\rm VE, Surr}$   &   \boldsymbol{$3.060{\rm{mV}}$} & \boldsymbol{$2.927{\rm{mV}}$}  & $5.324 \rm{mV} (4.906 \rm{mV})$   & $ 10.541 \rm{mV} ( 6.913\rm{mV})$  \\ 
         $\varepsilon_{\rm VE, Phy}$   &   \boldsymbol{$4.308 {\rm{mV}}$} & \boldsymbol{$3.016{\rm{mV}}$}  & $6.670 \rm{mV} ( 5.962 \rm{mV})$   & $ 9.802\rm{mV} ( 6.676\rm{mV})$  \\ 
       
        \hline

        \end{tabular}
\caption{Accuracy metric comparisons between CNPE and Bayesian calibration on the validation set of the Comparison dataset. Best results are highlighted in bold. For CNPE, median voltage errors are reported in parentheses. The noise model is $\mathcal{U}(-5.76 \rm{mV}, 5.76 \rm{mV})$.}
\label{tab:acc4}
\end{table}

\begin{table}[h!]
\centering
 \begin{tabular}{ |c|c|c|c|c| } 
        \hline
        Metric &  \multicolumn{2}{|c|}{Bayesian calibration} & \multicolumn{2}{|c|}{CNPE} \\
        \hline
         & Discharge $\rm{C}/2$ & Charge $4\rm{C}$ & Discharge $\rm{C}/2$ & Charge $4\rm{C}$  \\
        \hline
         $\varepsilon_{\rm PE}$   &  \boldsymbol{$9.69\%$} & $8.48\%$  & $9.88\%$ & \boldsymbol{$7.46\%$}\\
         $\varepsilon_{\rm PE, s}$   &  \boldsymbol{$12.87\%$} & $11.15\%$  & $13.08\%$ & \boldsymbol{$10.48\%$} \\ 
          $\varepsilon_{\rm CE}$   &  $10.12\%$  &  $6.45\%$ & \boldsymbol{$4.55\%$} &   \boldsymbol{$4.45\%$} \\ 
          $\varepsilon_{\rm VE, Surr}$   &   \boldsymbol{$5.843 {\rm{mV}}$} & \boldsymbol{$5.842{\rm{mV}}$}  &  $8.025 {\rm{mV}}  ( 7.716{\rm{mV}})     $  &  $ 14.358\rm{mV} ( 11.460 \rm{mV})$ \\ 
         
         $\varepsilon_{\rm VE, Phy}$   &   \boldsymbol{$6.594 {\rm{mV}}$} & \boldsymbol{$5.941{\rm{mV}}$}  &  $9.108 {\rm{mV}} ( 8.435{\rm{mV}})$  &  $ 13.513 {\rm{mV}} ( 10.860{\rm{mV}})$ \\ 

          %$\varepsilon_{\rm VE, Phy}$   &   $6.594 {\rm{mV}}$ & \boldsymbol{$5.941{\rm{mV}}$}  &  \boldsymbol{$6.217\rm{mV} ( 5.631{\rm{mV}})$}   &  $ 12.391\rm{mV} ( 10.407{\rm{mV}})$  \\ 
       
        \hline

        \end{tabular}
\caption{Accuracy metric comparisons between CNPE and Bayesian calibration on the validation set of the Comparison dataset. Best results are highlighted in bold. For CNPE, median voltage errors are reported in parentheses. The noise model is $\mathcal{U}(-11.52 \rm{mV}, 11.52 \rm{mV})$.}
\label{tab:acc8}
\end{table}

\section{Priors for the high-dimensional cases}
\label{app:priorhighdim}

Table~\ref{tab:priorhighdim} shows the prior ranges of the parameters added for the high-dimensional inference cases in Sec.~\ref{sec:tractability}. The first 10 parameters are added to the ones in Tab.~\ref{tab:priorcompds} to obtain the 16-dimensional case. The next 11 parameters are added to obtain the 27-dimensional case.

\begin{table}[h!]
\centering
\begin{tabular}{|c | c |} 
 \hline
 Scaled parameter & Ranges \\ [0.5ex] 
 \hline
 Solid anode active material volume fraction  $\varepsilon_{\rm{s, a, AM}}'$ & [0.7, 1.0]\\
 Solid anode diffusivity $D_{\rm s,a}'$ & [0.2, 2.0] \\
 Initial electrolyte \ce{Li^+} concentration $c_{\rm e,0}'$ & [0.5, 2.0] \\
 Electrolyte volume fraction in the anode $\varepsilon_{\rm{e, a}}'$ & [0.7, 1.0] \\
 Electrolyte volume fraction in the cathode $\varepsilon_{\rm{e, c}}'$ & [0.7, 1.0] \\
 Electrolyte volume fraction in the separator $\varepsilon_{\rm{e, s}}'$ & [0.7, 1.0] \\
 \ce{Li^+} diffusivity in the electrolyte $D_{\rm e}'$ & [0.2, 10.0] \\
 \ce{Li^+} transference number $t_0'$ & [0.5, 2.0] \\
 Electrolyte conductivity $\kappa'$ & [0.5, 2.0] \\
 Thermodynamic factor $\gamma'$ & [0.5, 2.0] \\ \hline
 Cell area $A'$ & [0.5, 1.5] \\
 Anode thickness $L_{\rm a}'$ & [0.5, 1.5] \\
 Separator thickness $L_{\rm s}'$ & [0.5, 1.5] \\
 Cathode thickness $L_{\rm c}'$ & [0.5, 1.5] \\
 Anode secondary particle radius $R_{\rm s, a}'$ & [0.5, 2.0] \\
 Cathode secondary particle radius $R_{\rm s, c}'$ & [0.5, 2.0] \\
 Liquid-phase Bruggeman factor $p_{\rm l}'$ & [0.5, 2.0] \\
 Anode solid-phase Bruggeman factor $p_{\rm s, a}'$ & [0.5, 2.0] \\
 Anode liquid-phase Bruggeman factor $p_{\rm l, a}'$ & [0.5, 2.0] \\
 Cathode solid-phase Bruggeman factor $p_{\rm s, c}'$ & [0.5, 2.0] \\
 Cathode liquid-phase Bruggeman factor $p_{\rm l, c}'$ & [0.5, 2.0] \\
 \hline
\end{tabular}
\caption{Ranges of the uniform priors used for the high-dimensional parameters inference in Sec.~\ref{sec:tractability}.}
\label{tab:priorhighdim}
\end{table}

\end{document}